\documentclass[lettersize,journal]{IEEEtran}
\usepackage{amsmath,amsfonts}
\usepackage{algorithmic}
\usepackage{algorithm}
\usepackage{array}
\usepackage[caption=false,font=normalsize,labelfont=sf,textfont=sf]{subfig}
\usepackage{textcomp}
\usepackage{stfloats}
\usepackage{url}
\usepackage{xcolor}

\usepackage{verbatim}
\usepackage{graphicx}
\usepackage{cite}
\usepackage{multirow}
\usepackage{hyperref}
\usepackage{array}
\usepackage{booktabs}
\begin{document}

\title{Accented Text-to-Speech Synthesis\\ with Limited Data}

\author{Xuehao Zhou,~\IEEEmembership{Student~Member,~IEEE,}
        Mingyang Zhang,~\IEEEmembership{Member,~IEEE,}
        Yi Zhou,~\IEEEmembership{Member,~IEEE,}\\
        Zhizheng Wu,~\IEEEmembership{Member,~IEEE,}
        Haizhou Li,~\IEEEmembership{Fellow,~IEEE,}
}


\maketitle

\begin{abstract}
This paper presents an accented text-to-speech (TTS) synthesis framework with limited training data. We study two aspects concerning accent rendering: phonetic (phoneme difference) and prosodic (pitch pattern and phoneme duration) variations. The proposed accented TTS framework consists of two models: an accented front-end for grapheme-to-phoneme (G2P) conversion and an accented acoustic model with integrated pitch and duration predictors for phoneme-to-Mel-spectrogram prediction. The accented front-end directly models the phonetic variation, while the accented acoustic model explicitly controls the prosodic variation. Specifically, both models are first pre-trained on a large amount of data, then only the accent-related layers are fine-tuned on a limited amount of data for the target accent. In the experiments, speech data of three English accents, i.e., General American English, Irish English, and British English Received Pronunciation, are used for pre-training. The pre-trained models are then fine-tuned with Scottish and General Australian English accents, respectively. Both objective and subjective evaluation results show that the accented TTS front-end fine-tuned with a small accented phonetic lexicon ($5k$ words) effectively handles the phonetic variation of accents, while the accented TTS acoustic model fine-tuned with a limited amount of accented speech data (approximately $3$ minutes) effectively improves the prosodic rendering including pitch and duration. The overall accent modeling contributes to improved speech quality and accent similarity.
\end{abstract}


\section{Introduction}
\IEEEPARstart{S}{tate}-of-the-art neural end-to-end (E2E) text-to-speech (TTS) systems \cite{wang2017tacotron,shen2018natural,li2019neural,ren2019fastspeech,ren2020fastspeech} can generate human-like natural speech from text input. In many applications, such as audiobooks, customer service, and movie dubbing, accented speech is required for an improved user experience. Unlike multi-speaker TTS~\cite{ping2017deep,chen2020multispeech,jia2018transfer} systems that produce the voice of a target speaker by conditioning on an utterance-level speaker representation, or emotional TTS \cite{whitehill2019multi, cai2021emotion, li2021controllable} systems that represent the specific emotion with an utterance-level emotion embedding, an accent is characterized at different levels of an utterance~\cite{munro2008foreign}. Therefore, it is challenging to clone speaker and accent simultaneously, especially when the available data for the target accent is limited. This study is motivated to address such problems in accented TTS synthesis.

A foreign accent occurs when a native speaker of the first language (L1) pronounces the second language (L2) due to different linguistic systems between L1 and L2 \cite{flege1995second,best1994emergence}. The pronunciation pattern of the segmental and suprasegmental structures affects the perception of a foreign accent \cite{behrman2014segmental, jugler2017perceptual}.
The attributes of a foreign accent can be mainly categorized into the variations of phoneme and prosody \cite{jugler2017perceptual}, both of which are essential components of accent representation \cite{loots2011automatic, de2006contribution}. 
 
Taking English as an example, the phonetic transcriptions of lexical words vary from accent to accent. Such differences can be described in two aspects~\cite{yan2003analysis}: (a) the difference between the phoneme sets, e.g., the vowel set /ax, ea, ia, ua, oh/ does not appear in both British and American English accents; (b) the difference between the phonetic transcriptions, e.g., the word `day' is transcribed as /dei/ in the American accent, while /dæi/ in the Australian accent. 
On the other hand, prosody attributes such as pitch~\cite{bolinger1958theory} and duration are shown to be effective in accent morphing~\cite{yan2003analysis}. In a study on Turkish~\cite{levi2005acoustic}, results suggest that fundamental frequency ($F0$) and duration play a critical role in the rendering of accents. Obvious differences are observed between lexically accented and unaccented syllables with respect to $F0$ peaks. In~\cite{winters2013perceived, rognoni2014testing}, $F0$ and duration are selectively transplanted to observe that the two prosodic cues significantly affect accentedness.

To build an accented TTS system, the research problem is how to effectively model the accent-specific phonetic and prosodic patterns. In this work, we consider that an accent can be characterized by three key attributes, namely, phonetic variation, pitch pattern, and phoneme duration. We propose a framework for accented TTS that consists of an accented front-end for grapheme-to-phoneme (G2P) conversion and an accented acoustic model for phoneme-to-Mel-spectrogram prediction. We then study ways to build an accented TTS system with a limited amount of training data. In particular, we leverage the pre-trained models to transfer the shared knowledge from the large-scale general data to an accented TTS system.

In this paper, we propose an accented front-end to model the accent-specific phonetic variation. The front-end \cite{deri2016grapheme} learns to convert input text into a phonetic sequence in the target accent. Meanwhile, we also integrate a pitch predictor and a duration predictor into the acoustic model to modulate the $F0$ and phoneme duration for accent rendering. 
The proposed techniques are validated in two scenarios. In the first scenario, we assume that we have only a small accented phonetic lexicon and explore how to handle the phonetic variation in the TTS front-end. 
In the second scenario, we further assume that we not only have a small accented phonetic lexicon, but also a small amount of accented speech data. With both limited lexicon and speech data, we explore how prosodic variation can be modeled alongside the phonetic variation. The main contributions of this work are summarized below:
\begin{itemize}
    \item We formulate the accent rendering problem by addressing three key attributes of speech, namely, phonetic variation, pitch pattern, and phoneme duration.  
    \item We address the phonetic variation problem by fine-tuning the TTS front-end with a small accented phonetic lexicon.
    \item We control the prosody of speech by incorporating $F0$ and duration predictors in the TTS acoustic model, which can be fine-tuned with limited accented speech data. The $F0$ and duration are further adopted as additional inputs to the attention-based decoder. 
\end{itemize}

The rest of the paper is organized as follows. In Section II, we introduce the related work to set the stage of this study. Our proposed method and framework are presented in Section III. The experimental setup and result analysis are shown in Section IV and Section V, respectively. Section VI concludes this paper.

\section{Related Work}
\subsection{Traditional Accented TTS Synthesis}
Accented speech synthesis for various languages has been investigated. In~\cite{waseem2014speech, kayte2015speech}, Indian accented speech synthesis is studied with Festvox, a unit selection tool. Kolluru el at. \cite{kolluru2014generating} propose a method to generate multi-accent pronunciations for individual speaker by building a space of accents. They convert the phoneme sequence in the canonical version to an accented sequence via joint sequence model interpolation. The interpolated weights as points specify the different accents within an accent space. This method effectively changes the pronunciation of phonemes across accents. However, the prosodic variation, e.g., pitch and phoneme duration, of accent attributes is not modeled explicitly. 

Anumanchipalli et al. \cite{anumanchipalli2013accent} present an intonation model that automatically predicts the appropriate intonation contours from text for a statistical parametric speech synthesis system. This intonation model takes the accent group, a sequence of intonation events, as the modeling unit. This work reveals that the unit of accent components, such as $F0$, has an extremely high correlation with the linguistic pattern from text, which motivates us to consider using $F0$ to control the prosodic variation in accent rendering.

\subsection{Neural E2E Accented TTS Synthesis}
With the advanced neural-network-based approaches, Abeysinghe et al. \cite{abeysinghe2022visualising} adopt the E2E TTS architecture for accented speech generation. They pre-train a native TTS model with an American speech corpus and then perform fine-tuning with a non-native speech corpus. They visualize and analyze the vowel space variation during the fine-tuning stage and claim that the vowel space of voice generated from the fine-tuned model is closer to the non-native speech than to the native speech. Liu et al. \cite{liu2022controllable} present a controllable accented TTS framework. They propose an accent intensity modeling method that quantifies the accent intensity for each sample and a consistency constraint loss subject to the total TTS training loss. Their results demonstrate the effectiveness of using intensity control on accent rendering.
Melechovsky et al. \cite{melechovsky2022accented} propose a controllable speech synthesis system based on a conditional variational autoencoder. Their proposed method is capable of generating a specific speaker's voice with an arbitrary target accent.

However, none of the above works address the phonetic variation of English accents, which is a significant aspect in accent morphing. Moreover, prior studies typically rely on a large amount of training data, which limits the scope of applications with low-resource settings. 

In this paper, we seek to model both phonetic and prosodic variations in accented TTS with limited training data. 
To model the prosodic information, Yasuda et al. \cite{yasuda2019investigation} develop an E2E Japanese speech synthesis system by capturing long-term dependencies related to pitch accents from text encoder with an additional self-attention layer. In addition, Yasuda et al. \cite{yasuda2022investigation} examine that the pre-trained PnG BERT can capture the information related to pitch accents for a Japanese TTS system. They perform fine-tuning PnG BERT together with a TTS system and a tone prediction task to force PnG BERT to enrich the pitch accent information. 
Therefore, we consider controlling prosodic information of accent rendering from the text encoder and extending it to accented speech generation in limited data scenarios.

\section{Methodology}

\begin{figure}
    \centering
    \includegraphics[width=0.48\textwidth]{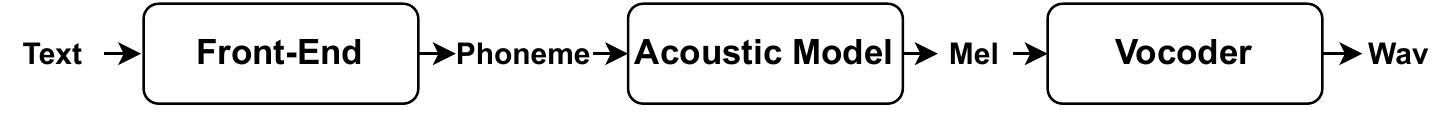}
    \caption{The system diagram of a typical TTS synthesis pipeline. Mel refers to the acoustic feature Mel-spectrogram, while Wav is the speech waveform.}
    \label{fig:tts_pipeline}
\end{figure}

A TTS system is generally a pipeline of three components, as shown in Fig. \ref{fig:tts_pipeline}: front-end, acoustic model, and vocoder. 
{The front-end plays a crucial role in a TTS system to provide the required phoneme-relevant linguistic knowledge \cite{parlikar2016festvox, pan2020unified, bansal2020improving}.}
We propose an accented TTS system that consists of an accented front-end and an accented acoustic model. The accented front-end model transcribes the text input into an accented phoneme representation, that handles the accented phonetic variation. 
{We take one of the state-of-the-art E2E TTS systems, Tacotron 2 \cite{shen2018natural} architecture, as the backbone of our accented acoustic model to predict the Mel-spectrogram with rich accented representation.}
The Parallel WaveGAN \cite{yamamoto2020parallel} is adopted as the neural vocoder to generate the speech waveform in the time-domain from the predicted Mel-spectrogram. 

We study two scenarios on the accented TTS framework, a) only a small accented phonetic lexicon is available for the target accent. b) both a small accented phonetic lexicon and limited accented speech samples are available for the target accent.

\begin{figure*}
    \centering
    \includegraphics[width=0.98\textwidth]{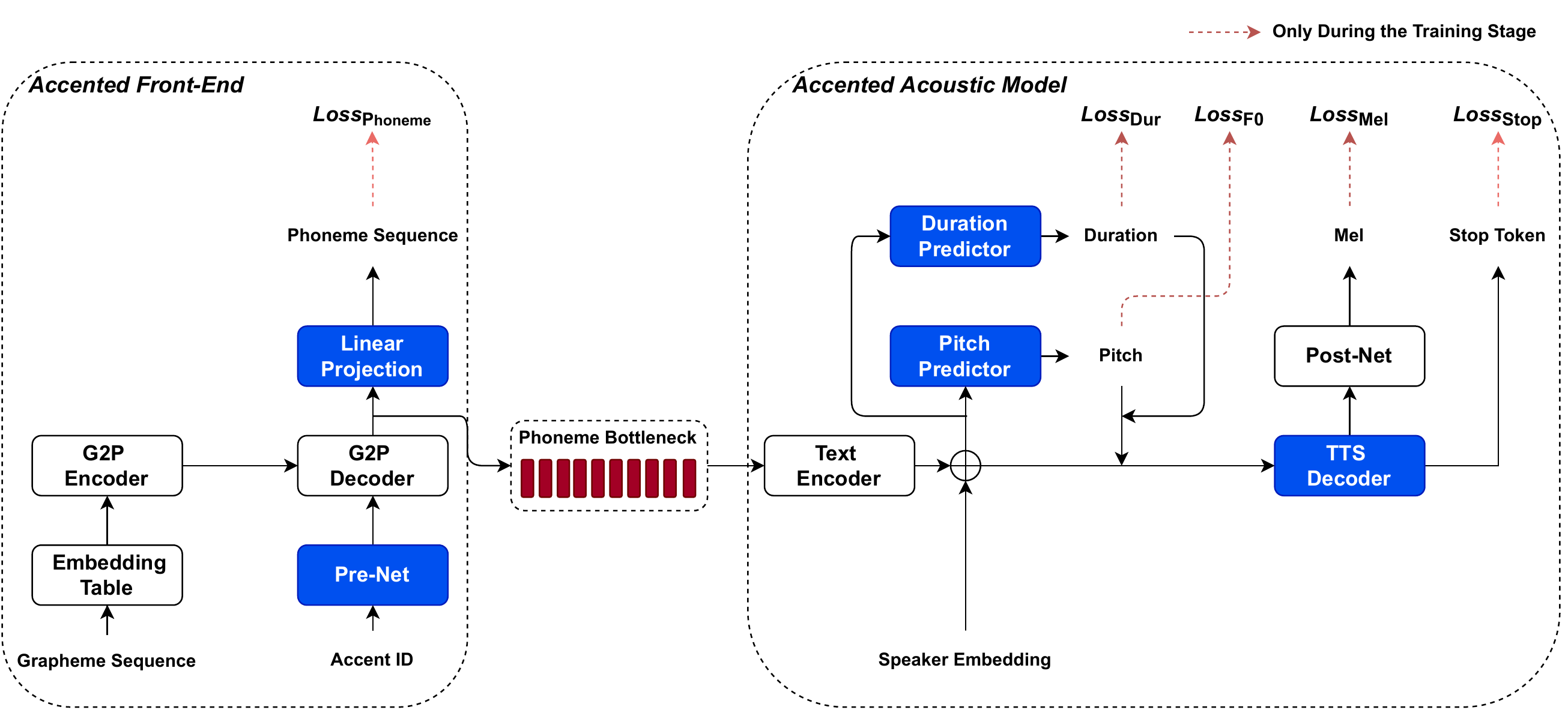}
    \caption{The block diagram of the proposed accented TTS framework. The left panel is the accented front-end for G2P conversion, while the right panel is the accented acoustic model for phoneme-to-Mel-spectrogram prediction. The phoneme bottleneck extracted from the G2P decoder output serves as the input of the acoustic model encoder. The front-end and acoustic model are pre-trained and then fine-tuned individually. All modules shown in this figure are involved during the pre-training stage. During the fine-tuning stage, only the parameters in the blue color modules are updated, while the other modules in the white color are fixed. The speaker encoder and neural vocoder are omitted for simplicity.}
    \label{fig:aTTS}
\end{figure*}

\subsection{Accented TTS Framework}
\subsubsection{Accented Front-End}
The Transformer-based model \cite{sevinj2019transformer} is well-known and effective for converting graphemes into phonemes.
However, it requires a large phonetic lexicon to build a standard Transformer-based G2P model, which is unrealistic for various English accents. It has been reported~\cite{dong2022neural, engelhart2021grapheme} that the performance of low-resource G2P models are improved by transferring shared knowledge from pre-trained models. Thus, we employ the technique of pre-training followed by fine-tuning to build an accented front-end with a limited size of accented phonetic lexicon. 

Inspired by the studies in multi-lingual and multi-speaker speech synthesis ~\cite{yu2016learning, gutkin2017uniform, kim2021gc}, we propose to pre-train a multi-accent G2P model with multiple English phonetic lexicons and fine-tune the accent-related layers with a small accented phonetic lexicon for the target accent. A multi-accent G2P model is similar to a multi-lingual one as both deal with the phonetic variation to different extents. An early study~\cite{sokolov2020neural} takes the language ID as an additional input for training a multi-lingual G2P model. We use an accent ID instead to pre-train a multi-accent G2P model in this work.

The left panel of Fig. \ref{fig:aTTS} is the multi-accent front-end for G2P conversion.   
The G2P model takes both an English grapheme sequence and an accent ID as the input and generates a phoneme sequence as the output. In practice, the grapheme sequence is converted to an embedding sequence via an embedding table and then encoded into a sequence of latent textual representations via a stack of Transformer encoders. The sequence of latent representations is of the same length as that of the input grapheme sequence.

For each accent, we have an accented phonetic lexicon. An English grapheme sequence may have several phonetic transcriptions, one for each accent. To predict the accented phoneme sequence, we use an accent ID as an additional input to the decoder. The phoneme sequence and accent ID are converted to the continuous space representation via a phoneme embedding table and an accent embedding table in the pre-net, respectively. The accent embedding is a vector that applies to the entire phoneme sequence. Specifically, the accent embedding is duplicated at the phoneme level and concatenated with the phoneme embedding. The combination of phoneme embedding and accent embedding serves as the input of the attention-based decoder.

A stack of Transformer decoders takes the encoded latent textual representation along with the accent embedding to generate an output sequence in an auto-regressive manner. The predicted phoneme of the previous step is taken as the decoder input at the current step.
The G2P decoder takes the ground truth phoneme sequence as input in a teacher-forcing manner during training. At run-time, the decoder takes the predicted phoneme sequence as the contextual input. Finally, the linear projection layer with a softmax function converts the output sequence to the discrete phoneme sequence. The loss function is defined as the cross-entropy between the generated phoneme sequence and ground truth.

\subsubsection{Accented Acoustic Model} 
The right panel of Fig. \ref{fig:aTTS} is a variant of TTS encoder-decoder acoustic model. In a standard TTS system, an embedding table maps a phoneme to its embedding vector in a continuous space. However, some phonemes in one accent may not exist in another accent. To generalize an accented front-end trained for one accent to another, we propose to make use of the embedding vector from the G2P decoder output, i.e., the phoneme bottleneck, instead of the discrete phoneme ID as the input to the TTS acoustic model. In this way, we hope to handle the pronunciation of unseen phonemes, even when the speech training data for the target accent is not available.

The proposed acoustic model also follows the strategy of pre-training and fine-tuning. It adopts the Tacotron 2 architecture, a data-driven model that represents the state-of-the-art performance in TTS. However, Tacotron 2 does not explicitly model the prosodic attributes of accents, e.g., pitch pattern and phoneme duration. Therefore, we incorporate pitch and duration predictors into the TTS acoustic model to enable the model to predict accurate pitch and phoneme duration that are close to ground truth.

As shown in Fig. \ref{fig:aTTS}, the text encoder takes the phoneme bottleneck features as input to generate the hidden phoneme representation. We condition the multi-speaker TTS pre-training on a $d$-vector-based speaker embedding, which represents the speaker identity. Since the pitch pattern and phoneme duration are also affected by the individual speaker, the speaker embedding is added to the text encoder output before the pitch and duration predictors. 
%
%
Phoneme modeling handled by the text encoder is an essential part of accented speech representation. Different from FastSpeech 2 \cite{ren2020fastspeech}, in this work, we adopt both duration predictor and attention mechanism to provide more accurate phoneme boundary with limited speech data, and the pitch and duration predictors are incorporated with text encoder to predict the $F0$ and duration at the phoneme level. 
The $F0$ and phoneme duration are added with the TTS acoustic model via an embedding table and a dense layer, respectively. 
The attention-based decoder network takes the combination of text encoder output, speaker embedding, $F0$, and phoneme duration to predict Mel-spectrogram and the stop token label. 

During the training stage, the ground truth of the Mel-spectrogram, $F0$, and phoneme duration are used in a teacher-forcing manner, while the predicted ones are used at run-time. 
The mean squared error is adopted as the loss function for the prediction of Mel-spectrogram, $F0$, and phoneme duration, while the stop token loss function is defined as the binary cross-entropy.
The total loss function of the proposed TTS acoustic model is:
\begin{equation}
\begin{split}
   Loss_{Total} = \alpha Loss_{Mel} + \beta Loss_{Stop} + \\
    \gamma Loss_{F0} + \delta Loss_{Dur}
\label{eq:sys}
\end{split}
\end{equation}
where $Loss_{Mel}$ is the sum of the Mel-spectrogram reconstruction loss before and after the post-net.

\subsection{Accented TTS Training with Limited Data}
\subsubsection{Only a Small Accented Phonetic Lexicon}
When a small accented phonetic lexicon for the target accent is available, we only fine-tune the accent-related layers of the pre-trained multi-accent G2P model. The accent-related layers are defined as the layers related to phoneme generation.
{Similar to \cite{engelhart2021grapheme}, we use the pre-trained model as initialization and fix the grapheme embedding table, the G2P encoder, and the G2P decoder. In \cite{engelhart2021grapheme}, the phoneme embedding table and the linear projection layer are re-initialized in the fine-tuning stage, because the phoneme sets are different in their pre-training and fine-tuning stages. In contrast, we use the unified phoneme sets for all accents involved in pre-training and fine-tuning stages. Therefore, instead of the re-initialization, we fine-tune the following layers based on their pre-trained parameters: the accent embedding table and phoneme embedding table in the pre-net, and the linear projection layer.} The fine-tuned modules are drawn in blue color, as shown in the Accented Front-End model of Fig. \ref{fig:aTTS}, while the other modules are fixed and shared for all accents during the fine-tuning stage.

\subsubsection{Both a Small Accented Phonetic Lexicon and Limited Accented Speech Samples}
Besides fine-tuning the pre-trained G2P model, we further address the prosodic variation in the generated accented speech by utilizing the limited accented speech samples for the target accent. Specifically, we use the pre-trained TTS acoustic model as initialization and fine-tune the accent-related layers to modify the prosodic components. As shown in Fig. \ref{fig:aTTS}, the fine-tuned modules of the Accented Acoustic Model are drawn in blue color.

In a previous study \cite{zhang2020adadurian}, it has been reported that fine-tuning the text encoder with limited speech data could degrade the TTS performance due to the uneven distribution of linguistic information. 
Hence, we freeze the text encoder during the fine-tuning stage. To control the phoneme duration, we fine-tune the duration predictor and the attention layer. To alter the pitch pattern, the pitch predictor and decoder are fine-tuned. The post-net after the decoder can improve the generalization ability of reconstruction across a large number of Mel-spectrogram during the pre-training stage so that a pre-trained post-net is fixed and shared for all accents during the fine-tuning stage. All the loss functions in Equation \ref{eq:sys} are optimized during the fine-tuning stage.

\begin{table}
\caption{A summary of the database used for building the accented acoustic model.}
\centering
\begin{tabular}{p{1.7cm}<{\centering} | p{1.3cm}<{\centering} | p{2.05cm}<{\centering} p{2.05cm}<{\centering}}
\specialrule{0.05em}{0pt}{1pt}
 & Pre-Train & \multicolumn{2}{c}{Fine-Tune} \\
\specialrule{0.05em}{1pt}{1pt}
Database  & VCTK & CMU\_ARCTIC & Google\_TTS\_API \\  
\specialrule{0.05em}{1pt}{1pt}
\# of Accent & 3 & 1 & 1 \\
\# of Speaker & 45 & 1 & 1 \\ 
\# of Utterance & 16,600 & 50 - 300 & 50 - 300 \\ 
Total duration  & 9.66 hours & 2.63 - 14.96 mins & 3.10 - 18.29 mins \\
\specialrule{0.05em}{1pt}{0pt}
\end{tabular}
\label{tab:tts_data}
\end{table}

\section{Experimental Setup}
We conduct experiments to validate the effectiveness of the proposed accented front-end model and accented acoustic model on accent rendering. Both models are first pre-trained on a large amount of data and then fine-tuned on limited accented data.

\begin{table}
\caption{A summary of model architectures and network configurations for the proposed accented TTS framework that consists of an accented G2P model and an accented TTS acoustic model. FC-x denotes a fully connected layer with x units. Conv1D-k-c means 1-D convolution with Width k and c Output Channels. LN denotes Layer Normalization, and BN indicates Batch Normalization.}
\centering
\begin{tabular}{p{2cm}<{\centering}|p{6cm}}
\specialrule{0.05em}{0pt}{1pt}
\multirow{2}*{G2P\_Embedding} & Grapheme embedding table: 256-D \\
~                             & Positional encoder: 256-D → Dropout(0.1) \\
\specialrule{0.05em}{1pt}{1pt}
\multirow{5}*{G2P\_Encoder} & Encoder stack number: 3 \\
~                           & Multi-head attention: 8 heads of 256-D \\
~                           & \qquad \qquad \qquad → Dropout(0.1) → LN \\ 
~                           & Feedforward: FC-512-ReLU → Dropout(0.1) \\
~                           & \qquad \qquad \qquad → FC-256 → Dropout(0.1) → LN \\ 
\specialrule{0.05em}{1pt}{1pt}
\multirow{2}*{G2P\_Attention} & Multi-head attention: 8 heads of 256-D \\
~                             & \qquad \qquad \qquad → Dropout(0.1) → LN \\ 
\specialrule{0.05em}{1pt}{1pt}
\multirow{4}*{G2P\_Pre-Net} & Phoneme embedding table: 256-D \\ 
~                           & Accent embedding table: 32-D \\ 
~                           & Linear projection: FC-256 \\ 
~                           & Positional encoding: 256-D → Dropout(0.1) \\ 
\specialrule{0.05em}{1pt}{1pt}
\multirow{5}*{G2P\_Decoder} & Decoder stack number: 3 \\ 
~                           & Masked multi-head attention: 8 heads of 256-D \\ 
~                           & \qquad \qquad \qquad → Dropout(0.1) → LN \\ 
~                           & Feedforward: FC-512-ReLU → Dropout(0.1) \\
~                           & \qquad \qquad \qquad → FC-256 → Dropout(0.1) → LN \\ 
\specialrule{0.05em}{1pt}{1pt}
G2P\_Projection & FC-82 \\
\specialrule{0.05em}{1pt}{1pt}
\multirow{4}*{TTS\_Encoder} & Phone linear projection: FC-512-softsign \\
~                           & 3 layers of conv1d-5-512-BN1d \\ 
~                           & \qquad \qquad \qquad → Bi-LSTM-512 \\ 
~                           & Speaker linear projection: FC-512-softsign \\
\specialrule{0.05em}{1pt}{1pt}
\multirow{2}*{TTS\_Attention} & Location layer: conv1d-31-32 \\
~                             & \qquad \qquad \qquad → Linear projection: FC-128 \\ 
\specialrule{0.05em}{1pt}{1pt} 
\multirow{6}*{TTS\_Decoder} & Pre-net: FC-256-ReLU → Dropout(0.5) \\
~                           & \qquad \qquad \qquad → FC-256-ReLU → Dropout(0.5) \\ 
~                           & RNN: LSTM-1024 → Dropout(0.1) \\ 
~                           & \qquad \qquad \qquad → LSTM-1024 → Dropout(0.1) \\
~                           & Mel linear projection: FC-80 \\
~                           & Stop token linear projection: FC-1 \\
\specialrule{0.05em}{1pt}{1pt} 
\multirow{2}*{TTS\_Post-Net} & 4 layers of conv1d-5-512-BN1d \\
~                           & \qquad \qquad \qquad → conv1d-5-80-BN1d \\
\specialrule{0.05em}{1pt}{1pt}
\multirow{4}*{Pitch Predictor} & Conv1d-3-512-ReLU-LN → Dropout(0.5) \\ 
~                              & \qquad \quad → conv1d-3-256-ReLU-LN → Dropout(0.5) \\
~                              & \qquad \quad → Linear projection: FC-1 \\ 
~                              & Pitch embedding table: 512-D \\ 
\specialrule{0.05em}{1pt}{1pt}
\multirow{4}*{Duration Predictor} & Conv1d-3-512-ReLU-LN → Dropout(0.5) \\ 
~                              & \qquad \quad → conv1d-3-256-ReLU-LN → Dropout(0.5) \\
~                              & \qquad \quad → Linear projection: FC-1 \\ 
~                              & Duration projection: FC-512-ReLU → Dropout(0.5) \\ 
\specialrule{0.05em}{1pt}{1pt} 
\multirow{2}*{Speaker Encoder} &   3 layers of LSTM-768 \\ 
~  & \qquad \quad → Linear projection: FC-256 \\ 
\specialrule{0.05em}{1pt}{0pt} 
\end{tabular}
\label{tab:network}
\end{table}

\subsection{Accented Front-End}
We first pre-train a multi-accent G2P model with accented phonetic lexicons of General American English, Irish English, and British English Received Pronunciation. Then, we fine-tune the pre-trained G2P model with Scottish and General Australian English phonetic lexicons, respectively. All phonetic lexicons are obtained from the Unisyn Lexicon~\cite{fitt2001morphological} that has a unified phoneme symbol inventory. During the fine-tuning stage, we vary the lexicon size of $1 k$, $5 k$, $10 k$, and $40 k$ most frequent words, in addition to the full lexicon size of about $120k$ words to observe the effect. When training the G2P model, the input is one utterance rather than a single word as in~\cite{engelhart2021grapheme}. The text transcripts are selected from the LibriTTS ~\cite{zen2019libritts} corpus. The utterances that are too short or too long and the utterances with any out-of-vocabulary word according to the phonetic lexicons are removed from the dataset.

Two front-end models are implemented for comparison:
\begin{itemize}
\item \textbf{SA-G2P:} This is a pre-trained single-accent G2P model trained only with General American English. 
\item \textbf{MA-G2P:} This is a pre-trained multi-accent G2P model trained with multiple accented phonetic lexicons including General American English, Irish English, and British English Received Pronunciation. The accent ID is used as an additional input in the multi-accent G2P model.
\end{itemize}

The model architectures and network configurations of the Transformer-based G2P model are shown in Table \ref{tab:network}.
Each G2P model is pre-trained for $100$ epochs and fine-tuned for another $50$ epochs for each experimental group. The model with the lowest validation loss among pre-training epochs is used as the initialization during the fine-tuning stage, and the model with the lowest validation loss among fine-tuning epochs is used for testing. The learning rate is $5e$-$4$ and the batch size is $128$ in both pre-training and fine-tuning stages. 

\subsection{Accented Acoustic Model}

We first pre-train a multi-speaker TTS acoustic model using the subset of the CSTR\_VCTK \cite{veaux2017cstr} speech corpus. To capture the shared knowledge across accents, we pre-train the model with speech data from multiple accents. 
We select the dataset with the same accents as we pre-train the G2P model, i.e., the accents of General American English, Irish English, and British English Received Pronunciation. Then, we fine-tune the pre-trained acoustic model with a Scottish speaker's data from CMU\_ARCTIC \cite{kominek2004cmu} corpus, and an Australian speaker's data generated by Google TTS API\footnote{\url{https://cloud.google.com/text-to-speech}}, respectively. We randomly select $50$, $100$, $200$, and $300$ utterances as the training data during the fine-tuning stage. The details of the involved database are summarized in Table \ref{tab:tts_data}. All speech signals are down-sampled to $16 k$ Hz and the silence at the beginning and end of each utterance is trimmed. We use the logarithmic scale $80$-dim Mel-spectrogram as the acoustic feature that is extracted with $12.5$ ms frame-shift and $50$ ms frame length. 

For the training of the phoneme-level pitch and duration predictors, we obtain the phoneme boundary by applying force-alignment using an automatic speech recognition (ASR) system. As a generic ASR model does not work well across accents, we train the accent-dependent ASR for force-alignment\footnote{\url{https://montreal-forced-aligner.readthedocs.io/en/latest}}. The phoneme duration is represented as the number of frames belonging to the same phoneme and transformed into a logarithmic scale.
The $F0$ is extracted using pyworld\footnote{\url{https://pypi.org/project/pyworld}} with $12.5$ ms frame shift. The linear interpolation method is adopted on the unvoiced frames of $F0$. The frame-level $F0$ is down-sampled to the phoneme-level according to the phoneme duration. Specifically, we take the average of $F0$ values on the frames belonging to the same phoneme. The phoneme-level $F0$ is then normalized to have zero mean and unit variance over the speech data for pre-training.
The $256$-dim speaker embedding is extracted from a pre-trained speaker encoder. The speaker encoder is trained with AISHELL-2 dataset \cite{du2018aishell}, following \cite{wan2018generalized}. 

The following six accented TTS systems are implemented for comparison:
\begin{itemize}
\item \textbf{Char-AM:} 
This is a multi-speaker TTS acoustic model that takes a character sequence as input.
\item \textbf{US\_G2P-AM:} 
This is an accented TTS framework that consists of an American G2P model and a multi-speaker TTS acoustic model. The American G2P model is part of the pre-trained multi-accent G2P model conditioned on the General American accent ID. 
\item \textbf{SCOT\_G2P-AM:} 
This is an accented TTS framework that consists of a Scottish G2P model and a multi-speaker TTS acoustic model. The Scottish G2P model is the pre-trained multi-accent G2P model fine-tuned with a Scottish phonetic lexicon of $5k$ words. 
\item \textbf{AU\_G2P-AM:} 
This is similar to SCOT\_G2P-AM except that the G2P model is fine-tuned with a General Australian phonetic lexicon of $5k$ words.  
\item \textbf{SCOT\_G2P-F0\_Dur\_AM:} 
This is an accented TTS framework that consists of a Scottish G2P model and a multi-speaker TTS acoustic model with integrated pitch and duration predictors. The Scottish G2P model is the pre-trained multi-accent G2P model fine-tuned with a Scottish phonetic lexicon of $5k$ words.
\item \textbf{AU\_G2P-F0\_Dur\_AM:} 
This is similar to SCOT\_G2P-F0\_Dur\_AM except that the G2P model is fine-tuned with a General Australian phonetic lexicon of $5k$ words.
\end{itemize}

In the following experiments, the TTS system followed with '-L' denotes using only a small accented phonetic lexicon, with '-S' denotes using only limited accented speech data, and with '-LS' denotes using both of them.

The model architectures and network configurations of the Tacotron 2-based acoustic model are summarized in Table \ref{tab:network}. Each of the TTS acoustic models is pre-trained for $800$ epochs and fine-tuned for another $100$ epochs for each experimental group using Adam optimizer \cite{kingma2014adam}. All weights $\alpha$, $\beta$, $\gamma$, and $\delta$ in Equation \ref{eq:sys} are set to $1$.
The learning rate is $1e$-$3$ and $1e$-$4$, and the batch size is $32$ and $8$ for the pre-training stage and fine-tuning stage, respectively.

\subsection{Waveform Generation}
We select Parallel WaveGAN \cite{yamamoto2020parallel} as the neural vocoder to reconstruct the time-domain speech waveform from the predicted Mel-spectrogram for all experiments owing to its capability of generating high-quality speech waveform with a rapid speed. It is pre-trained with the CSTR\_VCTK \cite{veaux2017cstr} corpus, with the same frame-shift and frame length as the TTS acoustic model.

\section{Experimental Analysis}
We conduct both objective and subjective evaluations. In this section, we report the experiments under two scenarios, a) only a small accented phonetic lexicon is available. b) both a small accented phonetic lexicon and limited accented speech samples are available. 

\subsection{Only a Small Accented Phonetic Lexicon}
In this scenario, we only fine-tune the pre-trained multi-accent G2P model, while the pre-trained multi-speaker TTS acoustic model is frozen. We evaluate the phonetic variation in the accent rendering in terms of the accuracy of the output phoneme of front-end model and the output speech of the complete system. 


\subsubsection{Objective Evaluation} 
We report the performance of the G2P model in terms of phoneme error rate (PER) and word error rate (WER). PER and WER indicate the Levenshtein distance between the predicted sequence and reference at the phoneme and word level, respectively. Lower PER and WER account for more accurate predicted phoneme sequences. To evaluate the phonetic variation of the output speech of TTS system, we calculate the Kullback-Leibler divergence \cite{joyce2011kullback} (KLD) of phonetic posteriorgram between the generated speech and reference speech to measure the phonetic distribution similarity. The lower value suggests a better performance. 


\begin{figure}
\includegraphics[width=0.45\textwidth]{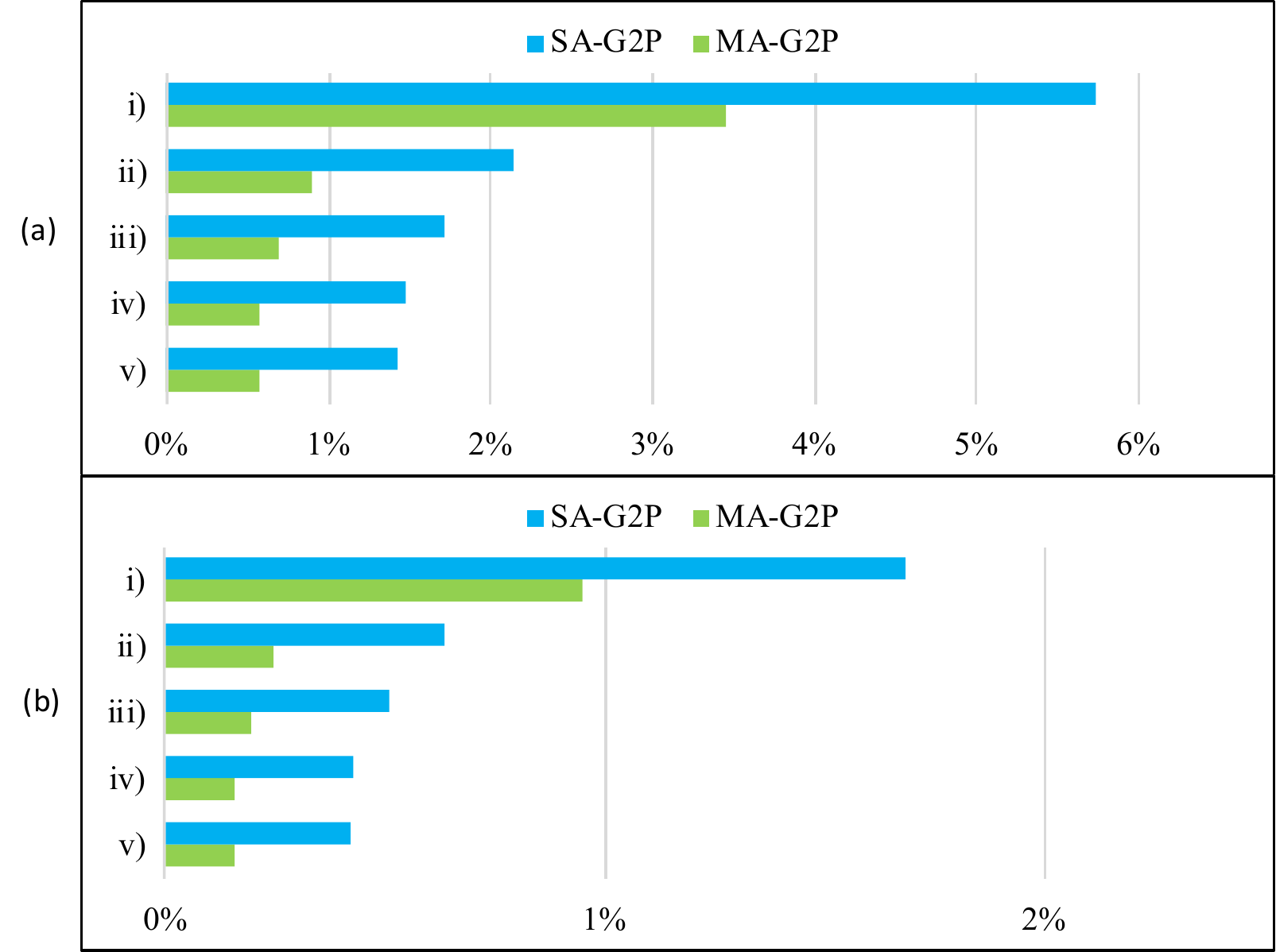}
\caption{The performance of both MA-G2P and SA-G2P fine-tuned with a Scottish phonetic lexicon of difference sizes in terms of WER and PER. (i) - (v) indicate the lexicon of $1k$, $5k$, $10k$, $40k$ words and full lexicon size, respectively. (a) WER (b) PER.}
\label{fig:g2p_scot}
\end{figure}

\begin{figure}
\includegraphics[width=0.45\textwidth]{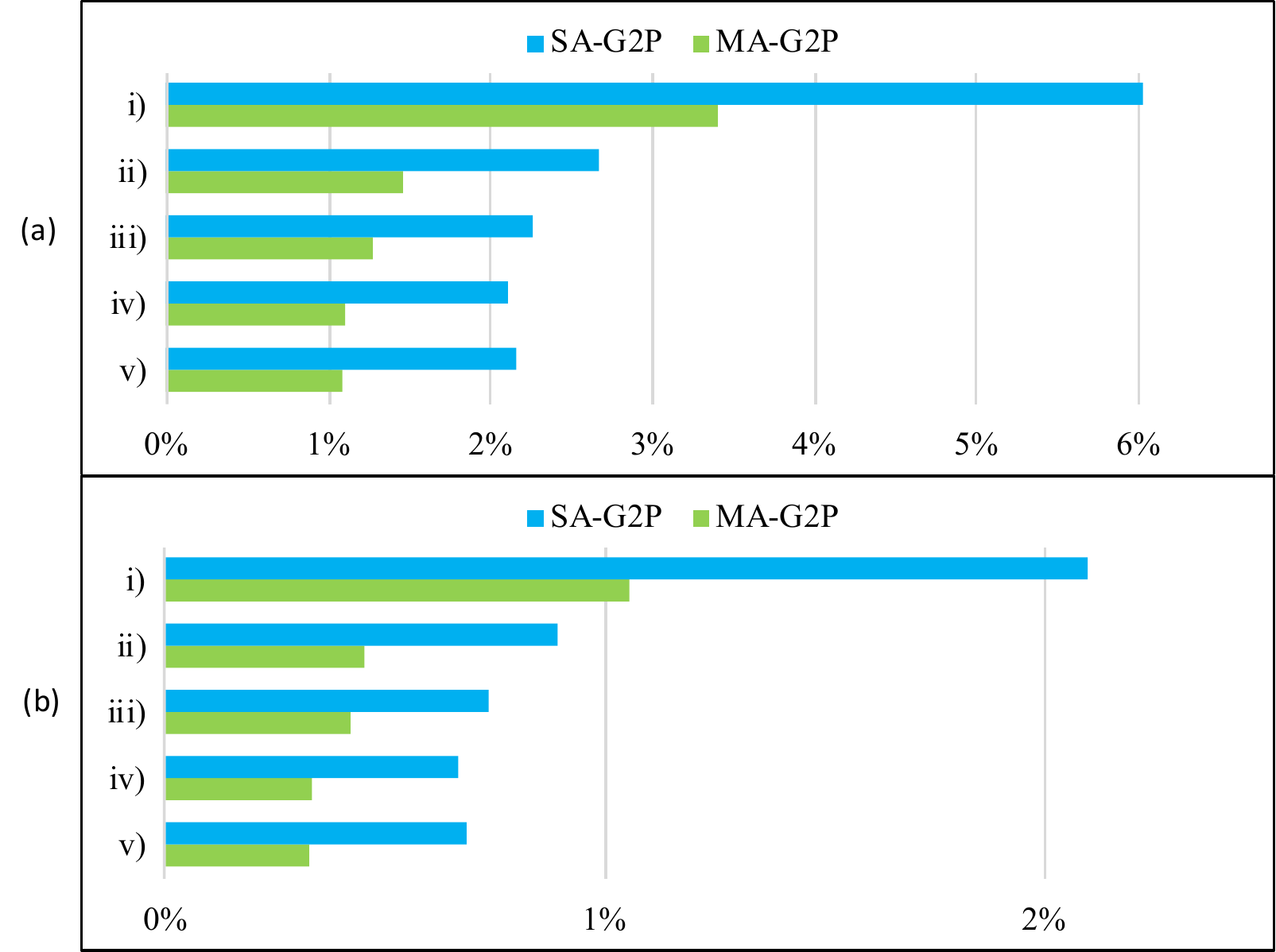}
\caption{The performance of both MA-G2P and SA-G2P fine-tuned with a General Australian phonetic lexicon of difference sizes in terms of WER and PER. (i) - (v) indicate the lexicon of $1k$, $5k$, $10k$, $40k$ words and full lexicon size, respectively. (a) WER (b) PER.}
\label{fig:g2p_au}
\end{figure}

\emph{i) Accuracy of Accented Front-End:} We compare two G2P models: accented G2P by fine-tuning MA-G2P and that by fine-tuning SA-G2P \cite{engelhart2021grapheme} on lexicons of different sizes.  The results on Scottish and Australian accents are illustrated in Fig. \ref{fig:g2p_scot} and Fig. \ref{fig:g2p_au}, respectively, and show that the fine-tuned MA-G2P outperforms the fine-tuned SA-G2P across all test cases. The results are also consistent between Scottish and Australian accents. 

It suggests that MA-G2P benefits from shared phonetic knowledge across accents. It is also observed that the Scottish G2P works better than the Australian counterpart, which can be explained by the fact that Scottish English accent is closer to Irish English and British English Received Pronunciation than the Australian English accent. 

As fine-tuning MA-G2P with a lexicon of $5k$ words achieves a reasonable performance (PER $0.25$\% and WER $0.90$\% on Scottish accent, PER $0.46$\% and WER $1.46$\% on Australian accent) that is comparable with full lexicon size (PER $0.16$\% and WER $0.58$\% on Scottish accent, PER $0.33$\% and WER $1.08$\% on Australian accent), we select the G2P fine-tuned with an accented phonetic lexicon of $5k$ words as our accented front-end in the limited data scenario for all subsequent experiments. 

\begin{table}
\caption{The comparison of phonetic transcriptions between Scottish / General Australian accent and General American accent}
\centering
\begin{tabular}{p{3.8cm}<{\centering} p{1.8cm}<{\centering}p{1.8cm}<{\centering}}
\specialrule{0.05em}{0pt}{1pt}
 & Scottish & Australian\\
\specialrule{0.05em}{1pt}{1pt}
Shared Word (\%) & 26.62 & 22.22 \\
Accented Word (\%) & 73.38 & 77.78 \\ 
Vowel Variation (\%) & 24.07 & 41.16 \\ 
Consonant Variation (\%) & 22.04 & 19.66 \\ 
\specialrule{0.05em}{1pt}{0pt}
\end{tabular}
\label{tab:lexicon_compare}
\end{table}

\begin{table}
\caption{KL divergence (KLD) of vowel distribution on the accented words for three comparative TTS systems on Scottish accent}
\centering
\begin{tabular}{p{2.5cm}<{\centering} p{2.5cm}<{\centering}}
\specialrule{0.05em}{0pt}{1pt}
 System & KLD\\
\specialrule{0.05em}{1pt}{1pt}
Char-AM             & 6.60\\ 
US\_G2P-AM-L         & 5.36 \\
SCOT\_G2P-AM-L        & \textbf{5.15} \\ 
\specialrule{0.05em}{1pt}{0pt}
\end{tabular}
\label{tab:KL_divergence}
\end{table}

\emph{ii) Phonetic Variation:} We perform statistical analysis on accented phonetic lexicons to understand the difference between Scottish / General Australian accent and General American accent. 
We count the percentage of the shared words whose phonetic transcriptions are the same between two lexicons and the accented words whose phonetic transcriptions are different between two lexicons. We further analyze the accented words in terms of vowel and consonant variations. Specifically, we calculate the rate of vowel and consonant differences by the Levenshtein distance on the accented words between two lexicons, respectively. 
As shown in Table \ref{tab:lexicon_compare}, the vowel variation is more prominent than the consonant one. The same findings have been reported in prior studies~\cite{yan2003analysis, yan2007analysis} on accented speech corpora. 

We therefore evaluate accent rendering by focusing on the vowels on the accented words. We select 191 accented words with a total of 304 vowel variations from the test set on Scottish accent. We extract the phonetic posteriorgram from a pre-trained speaker-independent ASR acoustic model. The frame-level posteriorgram is then down-sampled to the phoneme-level according to the phoneme boundary obtained by force-alignment. We calculate the KLD of the phoneme-level posteriorgram to compare the vowel distribution similarity, as shown in Table \ref{tab:KL_divergence}. 
We can observe that the generated speech from SCOT\_G2P-AM-L has a closer vowel distribution to reference speech than that from US\_G2P-AM-L on Scottish accent. We also observe that both G2P-AM-L outperform Char-AM. 

From the above observations, we could claim the following two statements: (a) TTS system with phoneme input is more able to model phonetic variation accurately than that with character input. (b) The pre-trained G2P fine-tuned with a small accented phonetic lexicon effectively changes the phonetic variation of output speech of TTS system.

\begin{figure}
\includegraphics[width=0.48\textwidth]{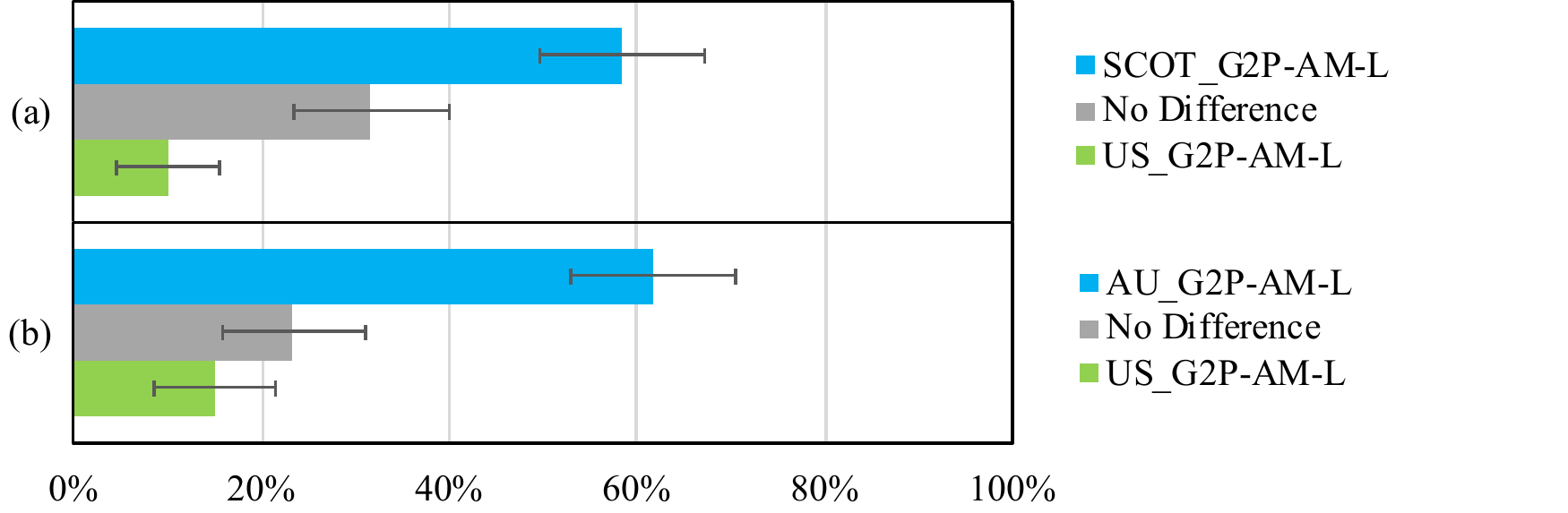}
\caption{XAB test results for accent pronunciation similarity of accented TTS systems with G2P pre-trained and fine-tuned with a small accented phonetic lexicon on both Scottish and Australian accents with 95\% confidence intervals. (a) Scottish accent. (b) Australian accent.}
\label{fig:gamvsedi}
\end{figure}


\subsubsection{Subjective Evaluation} 
The subjective evaluations are conducted through listening tests by human subjects. 20 listeners participate in each experimental set of the listening tests. All listeners are university students and staff members, who are proficient in English. \footnote{All speech samples are available at: \url{https://xuehao-marker.github.io/taslp_G2P-TTS/}} We conduct XAB preference test to assess the accent pronunciation similarity on the accented words on Scottish and Australian accents, respectively. In the XAB test, we label the accented words in each utterance in red color. The listeners are required to only pay attention to the labeled accented words and ignore other words. Each listener listens to the reference speech first and then chooses a more similar pronunciation to the reference speech from two different samples. Each listener listens to $18$ samples in a single experiment, thus in a total of $36$ ($18$ $\times$ $2$ (\# of accents) = $36$) samples. 

As shown in Fig. \ref{fig:gamvsedi}, US\_G2P-AM-L is compared with SCOT\_G2P-AM-L to show the effect of the accented G2P on Scottish accent. Also, US\_G2P-AM-L is compared with AU\_G2P-AM-L on Australian accent. It can be seen that the TTS system with fine-tuned G2P significantly achieves better performance than that with pre-trained US\_G2P in terms of accent pronunciation similarity, which is consistent on both Scottish and Australian accents. The results strongly support the idea of accented G2P modeling on addressing phonetic variation problem, that only depends on a small accented phonetic lexicon of $5k$ words without the need for accented speech samples during training.

\subsection{Both a Small Accented Phonetic Lexicon and Limited Accented Speech Samples}
After studying the effect of fine-tuning a pre-trained G2P model with a small accented phonetic lexicon, we further fine-tune the TTS acoustic model with a limited amount of accented speech data, to study the impact of the prosodic variation on accent rendering. We evaluate the performance on Scottish and Australian accents in terms of pitch pattern, phoneme duration, speech quality, and accent similarity. All experimental systems are tested on the same 100 unseen utterances.

\begin{table*}
\caption{The objective evaluation results in terms of pitch and duration of the comparative TTS systems fine-tuned with different sizes of accented speech data on Scottish and Australian accents.}
\centering
\begin{tabular}{p{1.2cm}<{\centering}  p{2.3cm}<{\centering}  p{3.5cm}<{\centering}  p{1.5cm}<{\centering} p{1.5cm}<{\centering} p{1.5cm}<{\centering}  p{1.8cm}<{\centering} p{1.45cm}<{\centering}}
\specialrule{0.05em}{0pt}{1pt}
Accent & \# of Utt / Duration & System & \multicolumn{3}{c}{Pitch} & \multicolumn{2}{c}{Duration}   \\
\specialrule{0.05em}{1pt}{1pt}
\multicolumn{3}{c}{}  & RMSE(Hz) & Correlation & U/V(\%) & Disturbance(frame) & RMSE(ms) \\
\specialrule{0.05em}{1pt}{1pt}
\multirow{15}*{Scottish} & \multirow{3}*{-} & Char-AM                      & 30.21 & 0.67 & 16.35 & 21.53 & 45.48 \\ 
~ & ~ & SCOT\_G2P-AM-L                                        & 35.33 & 0.67 & 15.27 & 19.18 & 51.49 \\
~ & ~ & SCOT\_G2P-$F0$\_Dur\_AM-L                               & 43.57 & 0.68 & 14.98 & 12.34 & 46.94 \\ 
\specialrule{0em}{0.5pt}{0.5pt}
\cline{2-8}
\specialrule{0em}{0.5pt}{0.5pt}
~ & \multirow{3}*{50 / 2.63 mins} & Char-AM-S                & 20.09 & 0.75 & 14.08 & 13.10 & 47.40 \\
~ & ~ & SCOT\_G2P-AM-LS                                        & 19.78  & 0.77 & 12.78 & 10.51  & 44.55 \\ 
~ & ~ & SCOT\_G2P-$F0$\_Dur\_AM-LS                               & \textbf{17.47} & \textbf{0.81} & \textbf{11.61} & \textbf{10.24} & \textbf{42.66} \\ 
\specialrule{0em}{0.5pt}{0.5pt}
\cline{2-8}
\specialrule{0em}{0.5pt}{0.5pt}
~ & \multirow{3}*{100 / 5.03 mins} & Char-AM-S               & 17.74 & 0.78 & 13.10 & 13.04 & 43.90 \\
~ & ~ & SCOT\_G2P-AM-LS                                        & 17.62 & 0.80 & 11.88 & 10.13 & 42.21 \\ 
~ & ~ & SCOT\_G2P-$F0$\_Dur\_AM-LS                               & \textbf{16.64} & \textbf{0.83} & \textbf{11.63} & \textbf{9.85} & \textbf{40.79} \\ 
\specialrule{0em}{0.5pt}{0.5pt}
\cline{2-8}
\specialrule{0em}{0.5pt}{0.5pt}
~ & \multirow{3}*{200 / 9.93 mins} & Char-AM-S               & 16.85 & 0.74 & 12.93 & 13.25 & 45.22 \\
~ & ~ & SCOT\_G2P-AM-LS                                        & 16.40 & 0.80 & 12.00 & 10.15 & \textbf{40.46} \\
~ & ~ & SCOT\_G2P-$F0$\_Dur\_AM-LS                               & \textbf{12.92} & \textbf{0.82} & \textbf{11.28} & \textbf{9.50} & 41.31 \\ 
\specialrule{0em}{0.5pt}{0.5pt}
\cline{2-8}
\specialrule{0em}{0.5pt}{0.5pt}
~ & \multirow{3}*{300 / 14.96 mins} & Char-AM-S              & 15.93 & 0.78 & 12.66 & 12.89 & 44.02 \\ 
~ & ~ & SCOT\_G2P-AM-LS                                        & 15.51 & 0.81 & 11.83 & 10.04 & 40.88 \\ 
~ & ~ & SCOT\_G2P-$F0$\_Dur\_AM-LS                               & \textbf{12.26} & \textbf{0.82} & \textbf{10.45} & \textbf{9.47} & \textbf{40.60} \\ 
\specialrule{0em}{0.5pt}{0.5pt}
\hline 
\specialrule{0em}{0.5pt}{0.5pt}
\multirow{15}*{Australian} & \multirow{3}*{-} & Char-AM                     & 40.88 & 0.37 & 19.04 & 12.85 & 40.11 \\ 
~ & ~ & AU\_G2P-AM-L                                        & 42.24 & 0.47 & 17.03 & 13.44 & 36.36 \\ 
~ & ~ & AU\_G2P-$F0$\_Dur\_AM-L                               & 40.48 & 0.54 & 17.73 & 9.97 & 41.01 \\ 
\specialrule{0em}{0.5pt}{0.5pt}
\cline{2-8} 
\specialrule{0em}{0.5pt}{0.5pt}
~ & \multirow{3}*{50 / 3.10 mins} & Char-AM-S              & 36.44 & 0.67 & 16.71 & 12.26 & 31.96 \\
~ & ~ & AU\_G2P-AM-LS                                        & 31.91 & 0.73 & 15.25 & 11.30 & 30.05 \\ 
~ & ~ & AU\_G2P-$F0$\_Dur\_AM-LS                               & \textbf{29.02} & \textbf{0.78} & \textbf{15.12} & \textbf{7.13} & \textbf{28.98} \\ 
\specialrule{0em}{0.5pt}{0.5pt}
\cline{2-8}
\specialrule{0em}{0.5pt}{0.5pt}
~ & \multirow{3}*{100 / 5.94 mins} & Char-AM-S             & 35.07 & 0.70 & 16.41 & 11.59 & 30.38 \\
~ & ~ & AU\_G2P-AM-LS                                        & 30.91 & 0.76 & 15.33 & 7.32 & 29.75 \\ 
~ & ~ & AU\_G2P-$F0$\_Dur\_AM-LS                               & \textbf{27.43} & \textbf{0.81} & \textbf{15.09} & \textbf{6.28} & \textbf{28.74} \\ 
\specialrule{0em}{0.5pt}{0.5pt}
\cline{2-8}
\specialrule{0em}{0.5pt}{0.5pt}
~ & \multirow{3}*{200 / 12.21 mins} & Char-AM-S            & 33.49 & 0.73 & 16.20 & 12.03 & 28.02 \\
~ & ~ & AU\_G2P-AM-LS                                        & 27.97 & 0.80 & 14.94 & 6.74 & 27.31 \\
~ & ~ & AU\_G2P-$F0$\_Dur\_AM-LS                               & \textbf{26.42} & \textbf{0.83} & \textbf{14.67} & \textbf{5.75} & \textbf{26.93} \\ 
\specialrule{0em}{0.5pt}{0.5pt}
\cline{2-8}
\specialrule{0em}{0.5pt}{0.5pt}
~ & \multirow{3}*{300 / 18.29 mins} & Char-AM-S            & 31.76 & 0.75 & 16.10 & 11.15 & 28.00 \\ 
~ & ~ & AU\_G2P-AM-LS                                        & 27.17 & 0.81 & 14.73 & 6.83 & 27.02 \\ 
~ & ~ & AU\_G2P-$F0$\_Dur\_AM-LS                               & \textbf{25.58} & \textbf{0.83} & \textbf{14.57} & \textbf{5.47} & \textbf{26.39} \\ 
\specialrule{0em}{0.5pt}{0.5pt}
\hline  
\specialrule{0em}{0.5pt}{0.5pt}
\multirow{3}*{Average} & \multirow{3}*{-} & Char-AM-S           & 25.92 & 0.74 & 14.77 & 12.42 & 37.36 \\ 
~ & ~ & G2P-AM-LS                                        & 23.41 & 0.78 & 13.59 & 9.13 & 35.28 \\ 
~ & ~ & G2P-$F0$\_Dur\_AM-LS                               & \textbf{20.97} & \textbf{0.82} & \textbf{13.05} & \textbf{7.96} & \textbf{34.55} \\ 
\specialrule{0.05em}{1pt}{0pt}
\end{tabular}
\label{tab:prosody_eval}
\end{table*}

\subsubsection{Objective Evaluation}
We calculate the $F0$ root mean squared error (RMSE), logarithmic scale $F0$ correlation coefficient \cite{kameoka2020convs2s}, and unvoiced/voiced (U/V) error rate to evaluate the pitch amplitude and trajectory trend similarity. We calculate the frame disturbance \cite{liu2021expressive} and phoneme duration RMSE to evaluate the accuracy of the predicted duration. Dynamic time warping (DTW) \cite{muller2007dynamic} is used to align the generated Mel-spectrogram and the reference. To further demonstrate the effectiveness of pitch and duration predictors on the fine-tuning with limited accented speech data, we fine-tune the TTS acoustic model with $50$, $100$, $200$, and $300$ utterances, respectively. The average objective test results are summarized in Table \ref{tab:prosody_eval}.

\begin{figure*}
\centering
{
\includegraphics[width=0.75\textwidth]{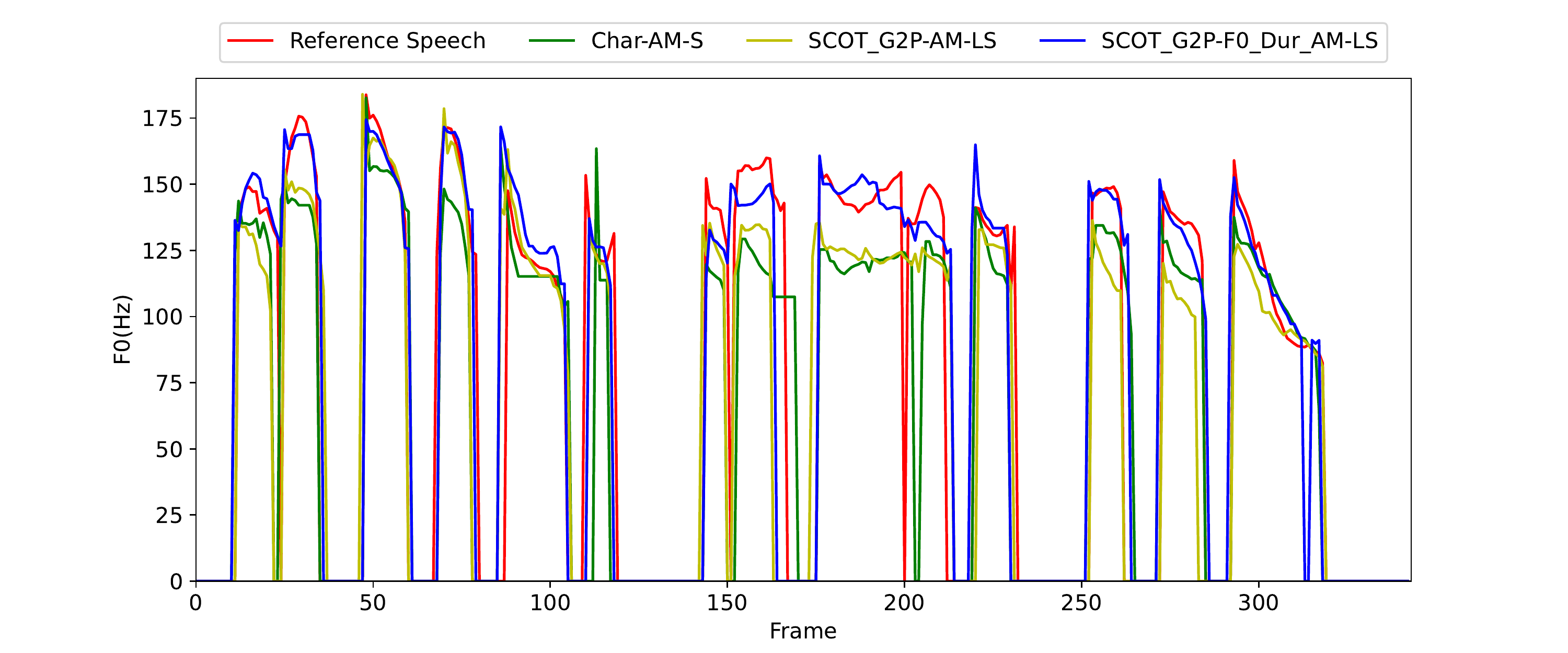}
}
\caption{The $F0$ contours on one Scottish utterance of the text transcription ``I just do appreciate it without being able to express my feelings". They are extracted from the reference speech and generated speech from three comparison systems fine-tuned with 300 accented utterances.}
\label{fig:f0}
\end{figure*}

\emph{i) Pitch:}
The $F0$ RMSE is defined as:
\begin{equation}
    RMSE = \sqrt{\frac{1}{N}\sum_{i=1}^{N}(x_i - \widehat{x}_i)^{2}}
    \label{equa:rmse}
\end{equation}
where $N$ is the total number of frames, $x_i$ and $\widehat{x}_i$ are the $F0$ values from the reference speech and generated speech at the $i^{th}$ frame. The lower $F0$ RMSE indicates the lower pitch amplitude error. Instead of log $F0$ RMSE \cite{ze2013statistical}, we calculate the original $F0$ RMSE over the both voiced frames between the aligned generated speech and reference speech.

The log-scale $F0$ correlation coefficient is defined as:
\begin{equation}
    C = \frac{\sum_{i=1}^{N}(y_i - \overline{y}_i)(\widehat{y}_i - \overline{\widehat{y}}_i)}{\sqrt{\sum_{i=1}^{N}(y_i - \overline{y}_i)^{2}}\sqrt{\sum_{i=1}^{N}(\widehat{y}_i - \overline{\widehat{y}}_i)^{2}}}
\end{equation}
where $\overline{y}=\frac{1}{N}\sum_{i=1}^{N}y$ and $\overline{\widehat{y}}=\frac{1}{N}\sum_{i=1}^{N}\widehat{y}$, $y_i$ and $\widehat{y}_i$ are the log-scale $F0$ values from the reference speech and generated speech at the $i^{th}$ frame and $N$ is the number of the calculation frames. We calculate the correlation coefficient only on both voiced frames of the $F0$ from aligned reference speech and generated speech. The correlation coefficient value is between the interval of $[-1, 1]$ and the closer to 1 indicates the higher pitch trajectory trend similarity.

We compare the U/V labels between the generated speech and reference speech in terms of U/V error rate, which is the ratio of the frame discrepancy count to the total number of frames. A lower U/V error rate indicates a better pitch reconstruction.

In Table \ref{tab:prosody_eval}, we make three observations, (a) All TTS systems fine-tuned with accented speech data outperform those without. This confirms the effectiveness of prosody modeling in the TTS acoustic model. It also suggests that the unseen $d$-vector-based speaker embedding is able to represent the speaker characteristic, while it lacks accent-related prosodic information. (b) From the four fine-tuning cases with different amounts of accented speech data, we note that the pitch performance of all G2P-AM-LS is better than that of Char-AM-S. This confirms the contribution of the G2P model to prosodic rendering. (c) We are glad to see that our proposed G2P-F0\_Dur\_AM-LS consistently achieves the lowest $F0$ RMSE, the highest correlation, and the lowest U/V error rate among the three compared systems in each fine-tuning case. This observation strongly demonstrates that the pitch predictor contributes to the more accurate pitch trajectory in the generated speech with only limited speech data for the target accent. The observations are consistent on both Scottish and Australian test cases.

We further visualize the pitch ($F0$) contours generated by the three comparative systems and the reference speech.
The $F0$ contours on one Scottish utterance are shown in Fig. \ref{fig:f0}. We observe that the $F0$ contour of SCOT\_G2P-F0\_Dur\_AM-LS is consistently the most similar one to the reference $F0$ across all systems, and collaborates the objective evaluation results in Table \ref{tab:prosody_eval}.

\emph{ii) Duration:}
The frame disturbance is defined as:
\begin{equation}
    Disturbance = \sqrt{\frac{1}{N}\sum_{i=1}^{N}(p_t - \widehat{p}_t)^{2}}
\end{equation}    
where $p_t$ and $\widehat{p}_t$ are the aligned path between the reference and generated speech at the $i^{th}$ frame, and $N$ is the number of the aligned frames. The lower disturbance value indicates the smaller duration distortion between the reference speech and generated speech. 

The definition of phoneme duration RMSE is similar to  Equation \ref{equa:rmse}, except that $x_i$ and $\widehat{x}_i$ here are the phoneme duration of the reference and generated speech for the $i^{th}$ phoneme and $N$ is the number of phonemes. A lower duration RMSE indicates better phoneme duration reconstruction.

In the last two columns of Table \ref{tab:prosody_eval}, we report the duration evaluation results on Scottish and Australian accents. We observe the same as in the pitch evaluation. (a) The duration reconstruction becomes more accurate when fine-tuning with limited accented speech data. (b) The overall performance of the three comparatively fine-tuned systems is ranked in a descending order: G2P-F0\_Dur\_AM-LS, G2P-AM-LS, and Char-AM-S. This observation is also consistent on Scottish and Australian accents. The results further confirm the contribution of the duration predictor to the duration reconstruction.

It is noted that the pitch and duration predictors highly rely on the small amount of accented speech training data. 

\subsubsection{Subjective Evaluation}
We evaluate the speech quality in terms of naturalness by mean opinion score (MOS) \cite{streijl2016mean} and further conduct XAB preference test and best-worst scaling (BWS) \cite{lee2008best} test to assess the accent similarity.

\begin{figure}
\includegraphics[width=0.48\textwidth]{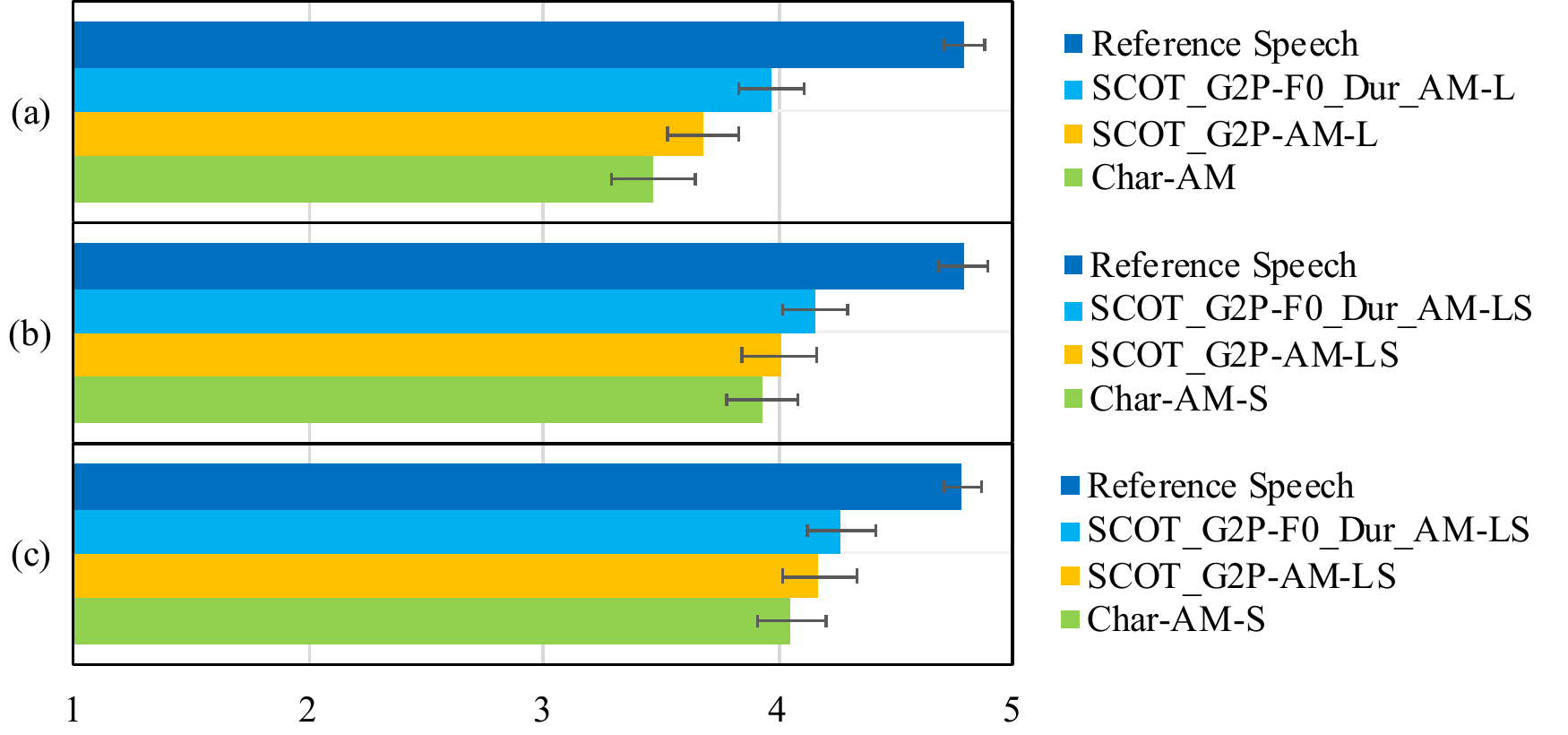}
\caption{MOS test results for naturalness of Char-AM, SCOT\_G2P-AM-L, SCOT\_G2P-F0\_Dur\_AM-L and Reference Speech on Scottish accent with 95\% confidence intervals. (a) Not fine-tuned with accented speech data. (b) Fine-tuned with 50 utterances. (c) Fine-tuned with 300 utterances.}
\label{fig:scot_mos}
\end{figure}

\begin{figure}
\includegraphics[width=0.48\textwidth]{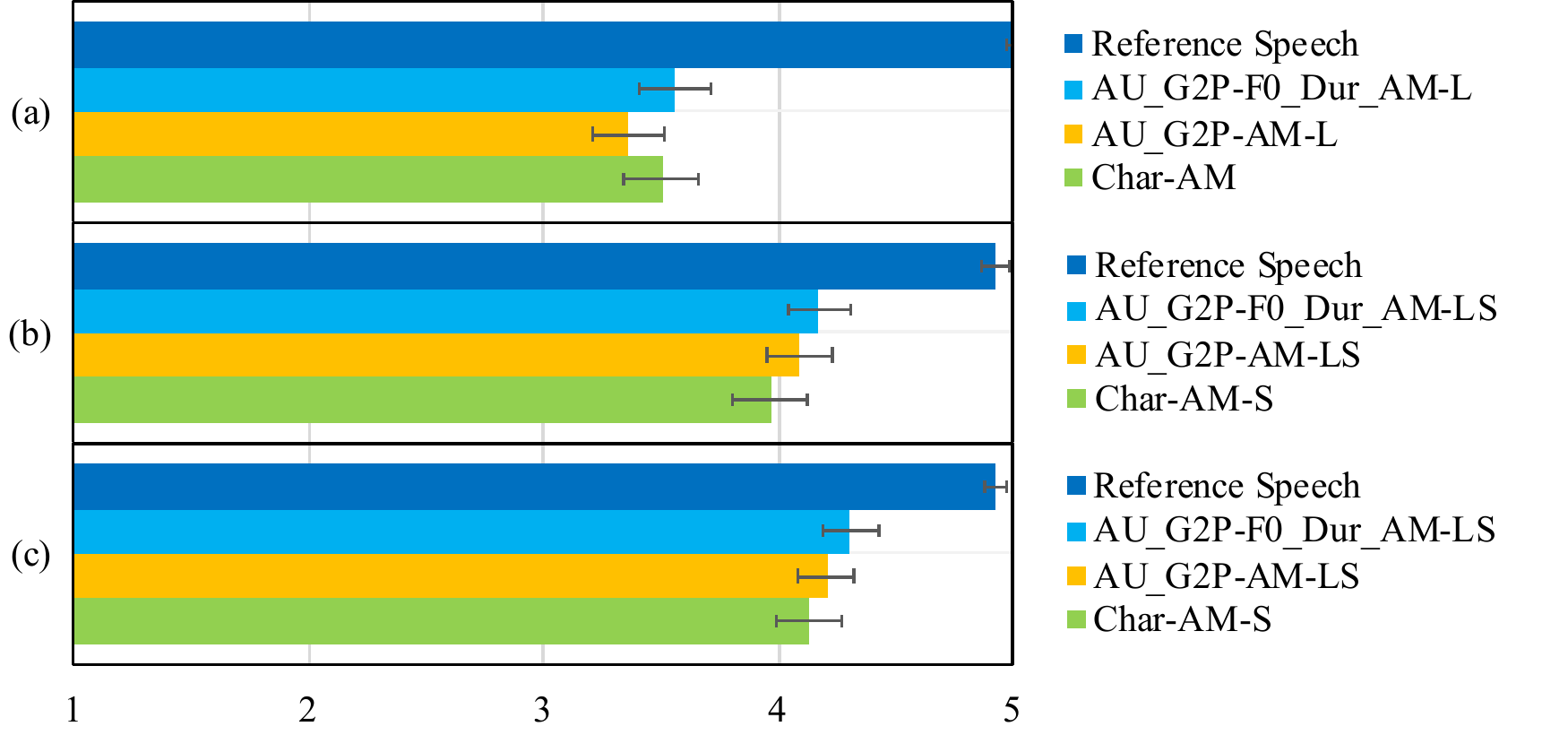}
\caption{MOS test results for naturalness of Char-AM, AU\_G2P-AM-L, AU\_G2P-F0\_Dur\_AM-L and Reference Speech on Australian accent with 95\% confidence intervals. (a) Not fine-tuned with accented speech data. (b) Fine-tuned with 50 utterances. (c) Fine-tuned with 300 utterances.}
\label{fig:aus_mos}
\end{figure}

\emph{i) Speech Quality:} 
In the MOS test, the listeners are asked to rate the speech naturalness of the provided samples on a $5$-point scale from $1$ to $5$. Each listener rates $24$ samples in a single experiment, thus in a total of $144$ ($24$ $\times$ $3$ (\# of fine-tuning cases) $\times$ $2$ (\# of accents) = $144$) samples. We evaluate the three comparative TTS systems and reference speech under three different scenarios, (a) Not fine-tuned with accented speech data, (b) Fine-tuned with $50$ accented speech utterances, (c) Fine-tuned with $300$ accented speech utterances, on both Scottish and Australian accents. The results are presented in Fig. \ref{fig:scot_mos} and Fig. \ref{fig:aus_mos}, respectively. 

In Fig. \ref{fig:scot_mos}, it is observed that SCOT\_G2P-F0\_Dur\_AM-L achieves the highest score in all three cases, followed by SCOT\_G2P-AM-L and Char-AM. This shows that phoneme input, pitch, and duration predictors help to generate higher quality and more natural speech. A similar conclusion can be drawn for Australian accent from Fig. \ref{fig:aus_mos}. However, we note that in Fig. \ref{fig:aus_mos} (a), the performance of AU\_G2P-F0\_Dur\_AM-L is not obviously better than that of Char-AM, and even AU\_G2P-AM-L obtains the slightly lower score than Char-AM. While in Fig. \ref{fig:scot_mos} (a) the performance of SCOT\_G2P-AM-L is much better than that of Char-AM. We suspect that the phoneme representation of the Australian accent is more distinctive than the Scottish accent when compared with the pre-training data of General American English, Irish English, and British English Received Pronunciation. The unseen phonemes in Australian accent adversely affect the speech quality and naturalness to some extent. This may be also the reason that both G2P-AM-L in Fig. \ref{fig:scot_mos} (a) achieve better performance than those two in Fig. \ref{fig:aus_mos} (a).

\begin{figure}
\includegraphics[width=0.48\textwidth]{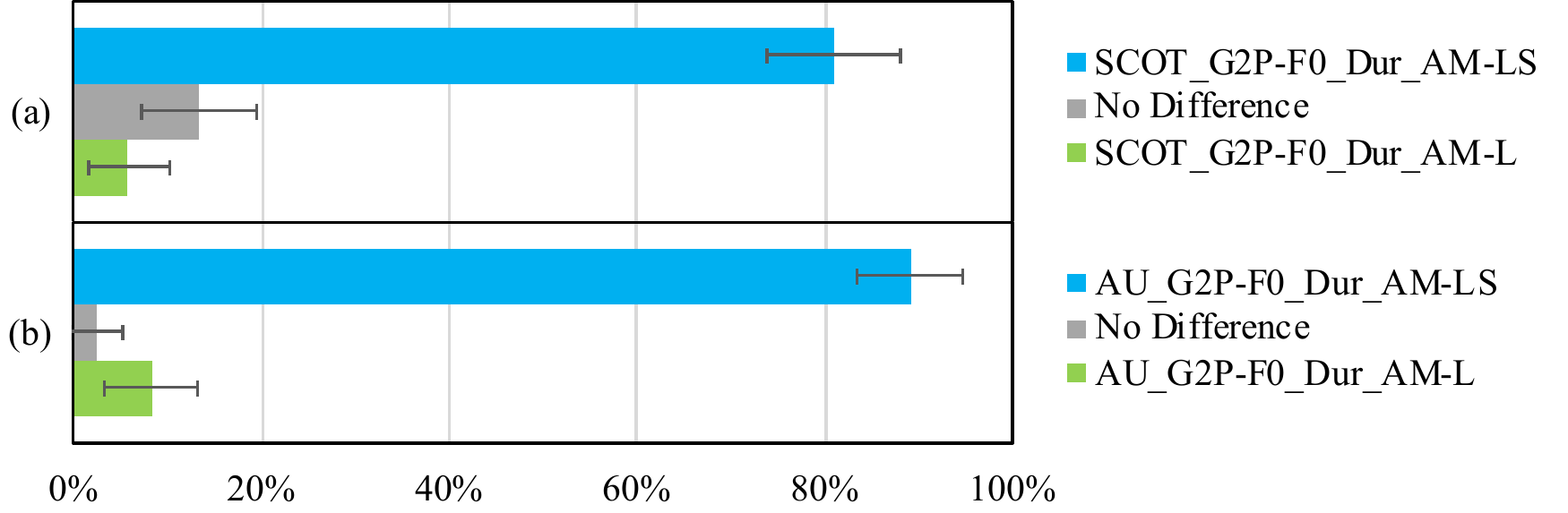}
\caption{XAB test results for accent similarity of G2P-F0\_Dur\_AM-L and G2P-F0\_Dur\_AM-LS on both Scottish and Australian accents with 95\% confidence intervals. (a) Scottish accent. (b) Australian accent.}
\label{fig:zsvs50}
\end{figure}

\begin{figure}
\includegraphics[width=0.48\textwidth]{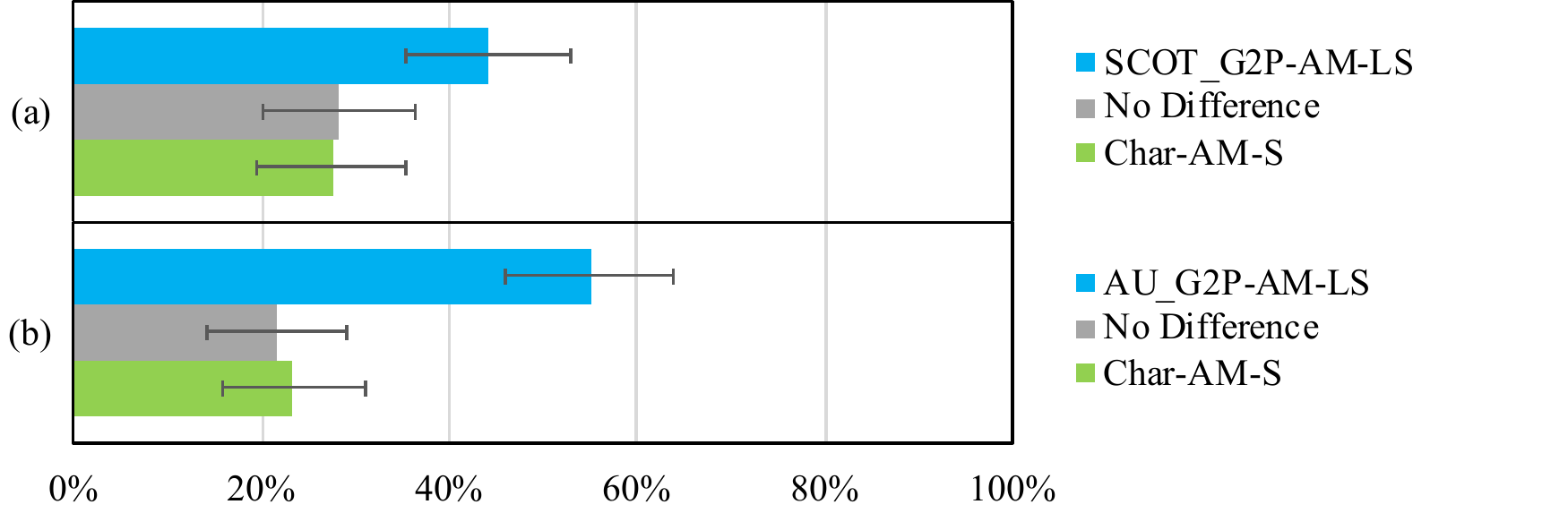}
\caption{XAB test results for accent similarity of Char-AM-S and G2P-AM-LS on both Scottish and Australian accents with 95\% confidence intervals. (a) Scottish accent. (b) Australian accent.}
\label{fig:chvspe}
\end{figure}

\begin{figure}
\includegraphics[width=0.48\textwidth]{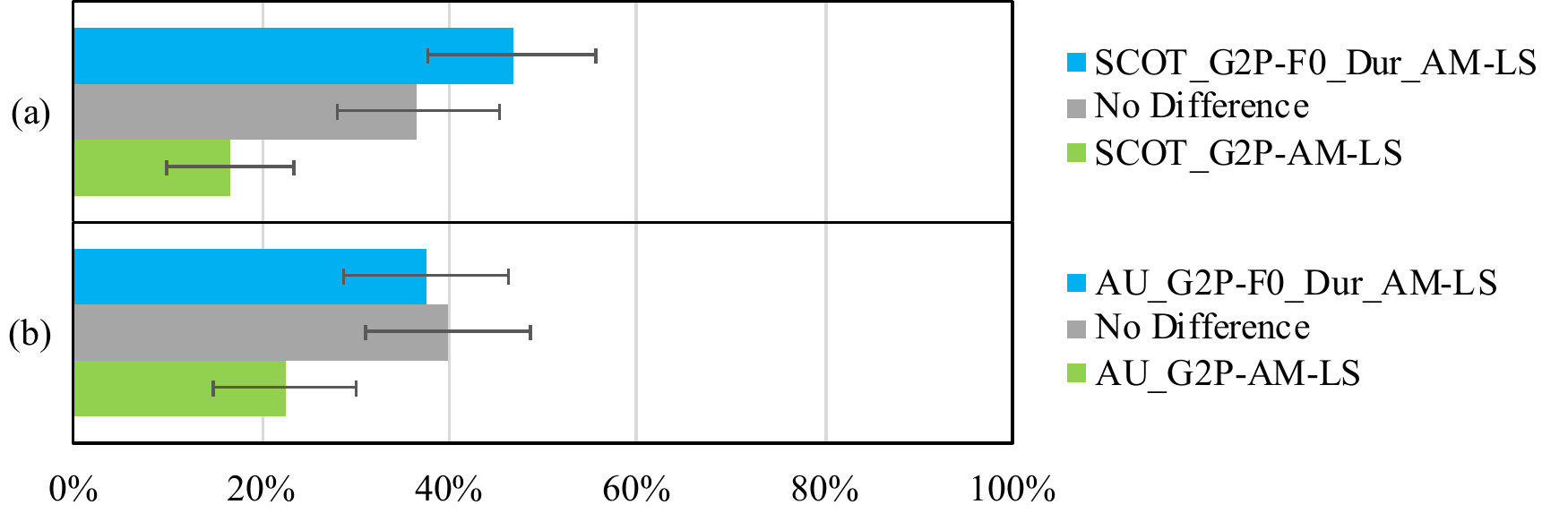}
\caption{XAB test results for accent similarity of G2P-AM-LS and G2P\-F0\_Dur\_AM-LS on both Scottish and Australian accents with 95\% confidence intervals. (a) Scottish accent. (b) Australian accent.}
\label{fig:pevsf0}
\end{figure}

\emph{ii) Accent Similarity:} In the both XAB preference and BWS tests, all TTS systems with '-LS' denote that the TTS acoustic models are fine-tuned with $50$ accented speech utterances. 

In the XAB preference test, the listeners are asked to listen to a reference speech first and then select the more similar sample to the reference speech from two different samples according to the accent similarity. Each listener listens to $18$ samples in a single experiment, thus in a total of $108$ ($18$ $\times$ $3$ (\# of comparison pairs) $\times$ $2$ (\# of accents) = $108$) samples. The results are shown in Fig. \ref{fig:zsvs50}-\ref{fig:pevsf0}. In Fig. \ref{fig:zsvs50}, it is apparent that the fine-tuned G2P-F0\_Dur\_AM-LS significantly outperforms the G2P-F0\_Dur\_AM-L. This shows that fine-tuning TTS acoustic model with limited accented speech data improves the prosodic rendering of generated accented speech. Fig. \ref{fig:chvspe} shows that the generated speech from TTS acoustic model with correct phoneme representation input has higher accent similarity than that with character input. This demonstrates that the phoneme-based encoder extracts the prosody-related textual information well, compared with the character-based encoder. In Fig. \ref{fig:pevsf0}, we are glad to see that after involving pitch and duration predictors into the TTS acoustic model, there is an obvious improvement in the perception of accent similarity in the generated speech. This strongly shows the effectiveness of the integrated pitch and duration predictors on accent rendering. All the reported conclusions are consistent on both Scottish and Australian accents.

\begin{table}[t]
    \centering 
    \caption{BWS test results for accent similarity of the four comparison systems on both Scottish and Australian accents.}
    \begin{tabular}{p{1cm}<{\centering}p{3.4cm}<{\centering}p{1.3cm}<{\centering}p{1.3cm}<{\centering}}
         \specialrule{0.05em}{0pt}{1pt}
         Accent & System & Best (\%) & Worst (\%) \\
         \specialrule{0.05em}{1pt}{1pt}
         \multirow{4}*{Scottish}    & SCOT\_G2P-F0\_Dur\_AM-L             & 7.5 & 78.33 \\ 
         ~                          & Char-AM-S                           & 13.33 & 13.33 \\ 
         ~                          & SCOT\_G2P-AM-LS                     & 25.83 & 5.00 \\ 
         ~                          & SCOT\_G2P-F0\_Dur\_AM-LS            & \textbf{53.33} & \textbf{3.33} \\ 
         \specialrule{0.05em}{1pt}{1pt}
          \multirow{4}*{Australian} & AU\_G2P-F0\_Dur\_AM-L               & 0.83 & 86.67 \\ 
         ~                          & Char-AM-S                           & 15.00 & 6.67 \\ 
         ~                          & AU\_G2P-AM-LS                       & 36.67 & 4.17 \\ 
         ~                          & AU\_G2P-F0\_Dur\_AM-LS              & \textbf{47.50} & \textbf{2.50} \\ 
         \specialrule{0.05em}{1pt}{1pt}
    \end{tabular}
    \label{tab:bws}
\end{table}

In the BWS test, we provide 5 samples of the same content to the listeners. They are asked to listen to the reference speech first and then choose the most and the least similar samples to the reference speech from four different samples according to the accent similarity. Each listener listens to $30$ samples in a single experiment, thus in a total of $60$ ($30$ $\times$ $2$ (\# of accents) = $60$) samples. The results on Scottish and Australian accents are shown in Table \ref{tab:bws}. It is observed that the proposed G2P-F0\_Dur\_AM-LS system obtains the highest best score and the lowest worst score among the four comparative systems. We also note that the G2P-F0\_Dur\_AM-L obtains significantly the lowest best score and the highest worst score. This indicates the effectiveness of fine-tuning with limited accented speech data on accent rendering. From the overall trend, the system performance on accent similarity can be ranked in an ascending order: G2P-F0\_Dur\_AM-L, Char-AM-S, G2P-AM-LS, and G2P-F0\_Dur\_AM-LS. The obtained conclusions are the same as those of the objective evaluations.

\section{Conclusion}
We propose and validate an accented TTS framework by addressing both phonetic and prosodic variations of accent rendering. The study shows that the phonetic variation can be modeled by an accented front-end with a small accented phonetic lexicon. Meanwhile, prosodic variation can be modeled by an accented TTS acoustic model with explicit pitch and duration control when a limited amount of accented speech data is available. The study also reveals that the accented front-end also contributes to accented prosodic rendering. The key finding of this work is that it is possible to effectively model a target accent with a limited amount of accented lexical and speech data. In the future work, we will explore a joint training of the accented front-end and acoustic model for accented TTS synthesis. 

\bibliographystyle{IEEEtran}
\bibliography{ref}

\begin{thebibliography}{10}
\providecommand{\url}[1]{#1}
\csname url@samestyle\endcsname
\providecommand{\newblock}{\relax}
\providecommand{\bibinfo}[2]{#2}
\providecommand{\BIBentrySTDinterwordspacing}{\spaceskip=0pt\relax}
\providecommand{\BIBentryALTinterwordstretchfactor}{4}
\providecommand{\BIBentryALTinterwordspacing}{\spaceskip=\fontdimen2\font plus
\BIBentryALTinterwordstretchfactor\fontdimen3\font minus
  \fontdimen4\font\relax}
\providecommand{\BIBforeignlanguage}[2]{{%
\expandafter\ifx\csname l@#1\endcsname\relax
\typeout{** WARNING: IEEEtran.bst: No hyphenation pattern has been}%
\typeout{** loaded for the language `#1'. Using the pattern for}%
\typeout{** the default language instead.}%
\else
\language=\csname l@#1\endcsname
\fi
#2}}
\providecommand{\BIBdecl}{\relax}
\BIBdecl

\bibitem{wang2017tacotron}
Y.~Wang, R.~Skerry-Ryan, D.~Stanton, Y.~Wu, R.~J. Weiss, N.~Jaitly, Z.~Yang,
  Y.~Xiao, Z.~Chen, S.~Bengio \emph{et~al.}, ``Tacotron: Towards end-to-end
  speech synthesis,'' \emph{arXiv preprint arXiv:1703.10135}, 2017.

\bibitem{shen2018natural}
J.~Shen, R.~Pang, R.~J. Weiss, M.~Schuster, N.~Jaitly, Z.~Yang, Z.~Chen,
  Y.~Zhang, Y.~Wang, R.~Skerrv-Ryan \emph{et~al.}, ``Natural tts synthesis by
  conditioning wavenet on mel spectrogram predictions,'' in \emph{2018 IEEE
  international conference on acoustics, speech and signal processing
  (ICASSP)}.\hskip 1em plus 0.5em minus 0.4em\relax IEEE, 2018, pp. 4779--4783.

\bibitem{li2019neural}
N.~Li, S.~Liu, Y.~Liu, S.~Zhao, and M.~Liu, ``Neural speech synthesis with
  transformer network,'' in \emph{Proceedings of the AAAI Conference on
  Artificial Intelligence}, vol.~33, no.~01, 2019, pp. 6706--6713.

\bibitem{ren2019fastspeech}
Y.~Ren, Y.~Ruan, X.~Tan, T.~Qin, S.~Zhao, Z.~Zhao, and T.-Y. Liu, ``Fastspeech:
  Fast, robust and controllable text to speech,'' \emph{Advances in Neural
  Information Processing Systems}, vol.~32, 2019.

\bibitem{ren2020fastspeech}
Y.~Ren, C.~Hu, X.~Tan, T.~Qin, S.~Zhao, Z.~Zhao, and T.-Y. Liu, ``Fastspeech 2:
  Fast and high-quality end-to-end text to speech,'' \emph{arXiv preprint
  arXiv:2006.04558}, 2020.

\bibitem{ping2017deep}
W.~Ping, K.~Peng, A.~Gibiansky, S.~O. Arik, A.~Kannan, S.~Narang, J.~Raiman,
  and J.~Miller, ``Deep voice 3: Scaling text-to-speech with convolutional
  sequence learning,'' \emph{arXiv preprint arXiv:1710.07654}, 2017.

\bibitem{chen2020multispeech}
M.~Chen, X.~Tan, Y.~Ren, J.~Xu, H.~Sun, S.~Zhao, T.~Qin, and T.-Y. Liu,
  ``Multispeech: Multi-speaker text to speech with transformer,'' \emph{arXiv
  preprint arXiv:2006.04664}, 2020.

\bibitem{jia2018transfer}
Y.~Jia, Y.~Zhang, R.~Weiss, Q.~Wang, J.~Shen, F.~Ren, P.~Nguyen, R.~Pang,
  I.~Lopez~Moreno, Y.~Wu \emph{et~al.}, ``Transfer learning from speaker
  verification to multispeaker text-to-speech synthesis,'' \emph{Advances in
  neural information processing systems}, vol.~31, 2018.

\bibitem{whitehill2019multi}
M.~Whitehill, S.~Ma, D.~McDuff, and Y.~Song, ``Multi-reference neural tts
  stylization with adversarial cycle consistency,'' \emph{arXiv preprint
  arXiv:1910.11958}, 2019.

\bibitem{cai2021emotion}
X.~Cai, D.~Dai, Z.~Wu, X.~Li, J.~Li, and H.~Meng, ``Emotion controllable speech
  synthesis using emotion-unlabeled dataset with the assistance of cross-domain
  speech emotion recognition,'' in \emph{ICASSP 2021-2021 IEEE International
  Conference on Acoustics, Speech and Signal Processing (ICASSP)}.\hskip 1em
  plus 0.5em minus 0.4em\relax IEEE, 2021, pp. 5734--5738.

\bibitem{li2021controllable}
T.~Li, S.~Yang, L.~Xue, and L.~Xie, ``Controllable emotion transfer for
  end-to-end speech synthesis,'' in \emph{2021 12th International Symposium on
  Chinese Spoken Language Processing (ISCSLP)}.\hskip 1em plus 0.5em minus
  0.4em\relax IEEE, 2021, pp. 1--5.

\bibitem{munro2008foreign}
M.~J. Munro, ``Foreign accent and speech intelligibility,'' \emph{Phonology and
  second language acquisition}, vol.~5, pp. 193--218, 2008.

\bibitem{flege1995second}
J.~E. Flege, ``Second language speech learning: Theory, findings, and
  problems,'' \emph{Speech perception and linguistic experience: Issues in
  cross-language research}, vol.~92, pp. 233--277, 1995.

\bibitem{best1994emergence}
C.~T. Best \emph{et~al.}, ``The emergence of native-language phonological
  influences in infants: A perceptual assimilation model,'' \emph{The
  development of speech perception: The transition from speech sounds to spoken
  words}, vol. 167, no. 224, pp. 233--277, 1994.

\bibitem{behrman2014segmental}
A.~Behrman, ``Segmental and prosodic approaches to accent management,''
  \emph{American Journal of Speech-Language Pathology}, vol.~23, no.~4, pp.
  546--561, 2014.

\bibitem{jugler2017perceptual}
J.~J{\"u}gler, F.~Zimmerer, J.~Trouvain, and B.~M{\"o}bius, ``The perceptual
  effect of l1 prosody transplantation on l2 speech: The case of french
  accented german,'' in \emph{Proceedings of Interspeech 2016. San Francisco,
  USA. 8--12 Sep, 2016}.\hskip 1em plus 0.5em minus 0.4em\relax International
  Speech Communication Association, 2017, pp. 67--71.

\bibitem{loots2011automatic}
L.~Loots and T.~Niesler, ``Automatic conversion between pronunciations of
  different english accents,'' \emph{Speech Communication}, vol.~53, no.~1, pp.
  75--84, 2011.

\bibitem{de2006contribution}
P.~B. De~Mare{\"u}il and B.~Vieru-Dimulescu, ``The contribution of prosody to
  the perception of foreign accent,'' \emph{Phonetica}, vol.~63, no.~4, pp.
  247--267, 2006.

\bibitem{yan2003analysis}
Q.~Yan, S.~Vaseghi, D.~Rentzos, C.-H. Ho, and E.~Turajlic, ``Analysis of
  acoustic correlates of british, australian and american accents,'' in
  \emph{2003 IEEE Workshop on Automatic Speech Recognition and Understanding
  (IEEE Cat. No. 03EX721)}.\hskip 1em plus 0.5em minus 0.4em\relax IEEE, 2003,
  pp. 345--350.

\bibitem{bolinger1958theory}
D.~L. Bolinger, ``A theory of pitch accent in english,'' \emph{Word}, vol.~14,
  no. 2-3, pp. 109--149, 1958.

\bibitem{levi2005acoustic}
S.~V. Levi, ``Acoustic correlates of lexical accent in turkish,'' \emph{Journal
  of the International Phonetic Association}, vol.~35, no.~1, pp. 73--97, 2005.

\bibitem{winters2013perceived}
S.~Winters and M.~G. O’Brien, ``Perceived accentedness and intelligibility:
  The relative contributions of f0 and duration,'' \emph{Speech Communication},
  vol.~55, no.~3, pp. 486--507, 2013.

\bibitem{rognoni2014testing}
L.~Rognoni and M.~G. Bus{\`a}, ``Testing the effects of segmental and
  suprasegmental phonetic cues in foreign accent rating: An experiment using
  prosody transplantation,'' in \emph{Proceeding of the International Symposium
  on the Acquisition of Second Language Speech, Concordia Working Papers in
  Applied Linguistics}, vol.~5, 2014, pp. 547--560.

\bibitem{deri2016grapheme}
A.~Deri and K.~Knight, ``Grapheme-to-phoneme models for (almost) any
  language,'' in \emph{Proceedings of the 54th Annual Meeting of the
  Association for Computational Linguistics (Volume 1: Long Papers)}, 2016, pp.
  399--408.

\bibitem{waseem2014speech}
M.~Waseem and C.~Sujatha, ``Speech synthesis system for indian accent using
  festvox,'' \emph{International journal of Scientific Engineering and
  Technology Research, ISSN}, pp. 2319--8885, 2014.

\bibitem{kayte2015speech}
S.~Kayte, M.~Mundada, and D.~C. Kayte, ``Speech synthesis system for marathi
  accent using festvox,'' \emph{International Journal of Computer
  Applications}, vol. 130, no.~6, pp. 38--42, 2015.

\bibitem{kolluru2014generating}
B.~Kolluru, V.~Wan, J.~Latorre, K.~Yanagisawa, and M.~J. Gales, ``Generating
  multiple-accent pronunciations for tts using joint sequence model
  interpolation,'' in \emph{Fifteenth Annual Conference of the International
  Speech Communication Association}, 2014.

\bibitem{anumanchipalli2013accent}
G.~K. Anumanchipalli, L.~C. Oliveira, and A.~W. Black, ``Accent group modeling
  for improved prosody in statistical parameteric speech synthesis,'' in
  \emph{2013 IEEE International Conference on Acoustics, Speech and Signal
  Processing}.\hskip 1em plus 0.5em minus 0.4em\relax IEEE, 2013, pp.
  6890--6894.

\bibitem{abeysinghe2022visualising}
B.~Abeysinghe, J.~James, C.~I. Watson, and F.~Marattukalam, ``Visualising model
  training via vowel space for text-to-speech systems,'' \emph{arXiv preprint
  arXiv:2208.09775}, 2022.

\bibitem{liu2022controllable}
R.~Liu, B.~Sisman, G.~Gao, and H.~Li, ``Controllable accented text-to-speech
  synthesis,'' \emph{arXiv preprint arXiv:2209.10804}, 2022.

\bibitem{melechovsky2022accented}
J.~Melechovsky, A.~Mehrish, B.~Sisman, and D.~Herremans, ``Accented
  text-to-speech synthesis with a conditional variational autoencoder,''
  \emph{arXiv preprint arXiv:2211.03316}, 2022.

\bibitem{yasuda2019investigation}
Y.~Yasuda, X.~Wang, S.~Takaki, and J.~Yamagishi, ``Investigation of enhanced
  tacotron text-to-speech synthesis systems with self-attention for pitch
  accent language,'' in \emph{ICASSP 2019-2019 IEEE International Conference on
  Acoustics, Speech and Signal Processing (ICASSP)}.\hskip 1em plus 0.5em minus
  0.4em\relax IEEE, 2019, pp. 6905--6909.

\bibitem{yasuda2022investigation}
Y.~Yasuda and T.~Toda, ``Investigation of japanese png bert language model in
  text-to-speech synthesis for pitch accent language,'' \emph{IEEE Journal of
  Selected Topics in Signal Processing}, 2022.

\bibitem{parlikar2016festvox}
A.~Parlikar, S.~Sitaram, A.~Wilkinson, and A.~W. Black, ``The festvox indic
  frontend for grapheme to phoneme conversion,'' in \emph{WILDRE: Workshop on
  Indian Language Data-Resources and Evaluation}, 2016.

\bibitem{pan2020unified}
J.~Pan, X.~Yin, Z.~Zhang, S.~Liu, Y.~Zhang, Z.~Ma, and Y.~Wang, ``A unified
  sequence-to-sequence front-end model for mandarin text-to-speech synthesis,''
  in \emph{ICASSP 2020-2020 IEEE International Conference on Acoustics, Speech
  and Signal Processing (ICASSP)}.\hskip 1em plus 0.5em minus 0.4em\relax IEEE,
  2020, pp. 6689--6693.

\bibitem{bansal2020improving}
S.~Bansal, A.~Mukherjee, S.~Satpal, and R.~Mehta, ``On improving code mixed
  speech synthesis with mixlingual grapheme-to-phoneme model.'' in
  \emph{INTERSPEECH}, 2020, pp. 2957--2961.

\bibitem{yamamoto2020parallel}
R.~Yamamoto, E.~Song, and J.-M. Kim, ``Parallel wavegan: A fast waveform
  generation model based on generative adversarial networks with
  multi-resolution spectrogram,'' in \emph{ICASSP 2020-2020 IEEE International
  Conference on Acoustics, Speech and Signal Processing (ICASSP)}.\hskip 1em
  plus 0.5em minus 0.4em\relax IEEE, 2020, pp. 6199--6203.

\bibitem{sevinj2019transformer}
Y.~Sevinj, N.~G{\'e}za, and G.-T. B{\'a}lint, ``Transformer based
  grapheme-to-phoneme conversion,'' \emph{Proceedings of Interspeech 2019}, pp.
  2095--2099, 2019.

\bibitem{dong2022neural}
L.~Dong, Z.-Q. Guo, C.-H. Tan, Y.-J. Hu, Y.~Jiang, and Z.-H. Ling, ``Neural
  grapheme-to-phoneme conversion with pre-trained grapheme models,'' in
  \emph{ICASSP 2022-2022 IEEE International Conference on Acoustics, Speech and
  Signal Processing (ICASSP)}.\hskip 1em plus 0.5em minus 0.4em\relax IEEE,
  2022, pp. 6202--6206.

\bibitem{engelhart2021grapheme}
E.~Engelhart, M.~Elyasi, and G.~Bharaj, ``Grapheme-to-phoneme transformer model
  for transfer learning dialects,'' \emph{arXiv preprint arXiv:2104.04091},
  2021.

\bibitem{yu2016learning}
Q.~Yu, P.~Liu, Z.~Wu, S.~K. Ang, H.~Meng, and L.~Cai, ``Learning cross-lingual
  information with multilingual blstm for speech synthesis of low-resource
  languages,'' in \emph{2016 IEEE International Conference on Acoustics, Speech
  and Signal Processing (ICASSP)}.\hskip 1em plus 0.5em minus 0.4em\relax IEEE,
  2016, pp. 5545--5549.

\bibitem{gutkin2017uniform}
A.~Gutkin, ``Uniform multilingual multi-speaker acoustic model for statistical
  parametric speech synthesis of low-resourced languages,'' 2017.

\bibitem{kim2021gc}
J.-H. Kim, S.-H. Lee, J.-H. Lee, H.-G. Jung, and S.-W. Lee, ``Gc-tts: Few-shot
  speaker adaptation with geometric constraints,'' in \emph{2021 IEEE
  International Conference on Systems, Man, and Cybernetics (SMC)}.\hskip 1em
  plus 0.5em minus 0.4em\relax IEEE, 2021, pp. 1172--1177.

\bibitem{sokolov2020neural}
A.~Sokolov, T.~Rohlin, and A.~Rastrow, ``Neural machine translation for
  multilingual grapheme-to-phoneme conversion,'' \emph{arXiv preprint
  arXiv:2006.14194}, 2020.

\bibitem{zhang2020adadurian}
Z.~Zhang, Q.~Tian, H.~Lu, L.-H. Chen, and S.~Liu, ``Adadurian: Few-shot
  adaptation for neural text-to-speech with durian,'' \emph{arXiv preprint
  arXiv:2005.05642}, 2020.

\bibitem{fitt2001morphological}
S.~Fitt, ``Morphological approaches for an english pronunciation lexicon,''
  \emph{International Speech Communication Association}, 2001.

\bibitem{zen2019libritts}
H.~Zen, V.~Dang, R.~Clark, Y.~Zhang, R.~J. Weiss, Y.~Jia, Z.~Chen, and Y.~Wu,
  ``Libritts: A corpus derived from librispeech for text-to-speech,''
  \emph{arXiv preprint arXiv:1904.02882}, 2019.

\bibitem{veaux2017cstr}
C.~Veaux, J.~Yamagishi, K.~MacDonald \emph{et~al.}, ``Cstr vctk corpus: English
  multi-speaker corpus for cstr voice cloning toolkit,'' \emph{University of
  Edinburgh. The Centre for Speech Technology Research (CSTR)}, 2017.

\bibitem{kominek2004cmu}
J.~Kominek and A.~W. Black, ``The cmu arctic speech databases,'' in \emph{Fifth
  ISCA workshop on speech synthesis}, 2004.

\bibitem{du2018aishell}
J.~Du, X.~Na, X.~Liu, and H.~Bu, ``Aishell-2: Transforming mandarin asr
  research into industrial scale,'' \emph{arXiv preprint arXiv:1808.10583},
  2018.

\bibitem{wan2018generalized}
L.~Wan, Q.~Wang, A.~Papir, and I.~L. Moreno, ``Generalized end-to-end loss for
  speaker verification,'' in \emph{2018 IEEE International Conference on
  Acoustics, Speech and Signal Processing (ICASSP)}.\hskip 1em plus 0.5em minus
  0.4em\relax IEEE, 2018, pp. 4879--4883.

\bibitem{kingma2014adam}
D.~P. Kingma and J.~Ba, ``Adam: A method for stochastic optimization,''
  \emph{arXiv preprint arXiv:1412.6980}, 2014.

\bibitem{joyce2011kullback}
J.~M. Joyce, ``Kullback-leibler divergence,'' in \emph{International
  encyclopedia of statistical science}.\hskip 1em plus 0.5em minus 0.4em\relax
  Springer, 2011, pp. 720--722.

\bibitem{yan2007analysis}
Q.~Yan, S.~Vaseghi, D.~Rentzos, and C.-H. Ho, ``Analysis and synthesis of
  formant spaces of british, australian, and american accents,'' \emph{IEEE
  Transactions on Audio, Speech, and Language Processing}, vol.~15, no.~2, pp.
  676--689, 2007.

\bibitem{kameoka2020convs2s}
H.~Kameoka, K.~Tanaka, D.~Kwa{\'s}ny, T.~Kaneko, and N.~Hojo, ``Convs2s-vc:
  Fully convolutional sequence-to-sequence voice conversion,'' \emph{IEEE/ACM
  Transactions on Audio, Speech, and Language Processing}, vol.~28, pp.
  1849--1863, 2020.

\bibitem{liu2021expressive}
R.~Liu, B.~Sisman, G.~Gao, and H.~Li, ``Expressive tts training with frame and
  style reconstruction loss,'' \emph{IEEE/ACM Transactions on Audio, Speech,
  and Language Processing}, vol.~29, pp. 1806--1818, 2021.

\bibitem{muller2007dynamic}
M.~M{\"u}ller, ``Dynamic time warping,'' \emph{Information retrieval for music
  and motion}, pp. 69--84, 2007.

\bibitem{ze2013statistical}
H.~Ze, A.~Senior, and M.~Schuster, ``Statistical parametric speech synthesis
  using deep neural networks,'' in \emph{2013 ieee international conference on
  acoustics, speech and signal processing}.\hskip 1em plus 0.5em minus
  0.4em\relax IEEE, 2013, pp. 7962--7966.

\bibitem{streijl2016mean}
R.~C. Streijl, S.~Winkler, and D.~S. Hands, ``Mean opinion score (mos)
  revisited: methods and applications, limitations and alternatives,''
  \emph{Multimedia Systems}, vol.~22, no.~2, pp. 213--227, 2016.

\bibitem{lee2008best}
J.~A. Lee, G.~Soutar, and J.~Louviere, ``The best--worst scaling approach: An
  alternative to schwartz's values survey,'' \emph{Journal of personality
  assessment}, vol.~90, no.~4, pp. 335--347, 2008.

\end{thebibliography}

 




\vfill

\end{document}